\documentclass{JHEP}
\usepackage{amsfonts}

\newcommand{\beq}{\begin{equation}} \newcommand{\eeq}{\end{equation}}
\newcommand{\beqa}{\begin{eqnarray}} \newcommand{\eeqa}{\end{eqnarray}}
\newcommand{\la}{\langle} \newcommand{\ra}{\rangle}

\newcommand{\ket}[1]{| #1 \rangle}
\newcommand{\bra}[1]{\langle #1 |}
\newcommand{\proj}[1]{\ket{#1}\bra{#1}}

\def\ot{\otimes}
\def\vp{\varphi}
\newcommand{\h}{\gamma_{ij}}
\newcommand{\hi}{\gamma_{i}}
\def\pplus{P_{i}^{xy}}
\def\12{\frac{1}{2}}

\newcommand{\proi}[1]{P_i(#1)}
\newcommand{\pai}[2]{P^{(1)}_{#2}(#1)}
\newcommand{\pbi}[2]{P^{(2)}_{#2}(#1)}
\newcommand{\A}{A(x)}
\newcommand{\B}{B(y)}
\newcommand{\Q}[1]{Q_{#1}}

\newcommand{\lone}[2]{L^{(1)}_{#2}(#1)}
\newcommand{\ltwo}[2]{L^{(2)}_{#2}(#1)}

\newcommand{\bQ}{{\bar Q}}
\newcommand{\gL}{M}

\def\s{\,\,\,}

\def\x{{x}}
\def\xv{{\bar x}}
\def\yv{{\bar y}}
\def\zv{{\bar z}}
\def\ug{e^{-iG\epsilon}}
\def\ugd{e^{iG\epsilon}}
\def\t{T}
\def\tr{tr}

\title{Fundamental destruction of information and conservation laws}
\abstract{
Theories which have fundamental information destruction or decoherence are motivated by
the black hole information paradox where one appears to have pure states evolving into
mixed states.
However such theories
have either violated conservation laws, or are highly non-local.  Here, we show
that the tension between conservation laws and locality can be circumvented by
constructing a relational theory of information destruction.  
In terms of conservation laws, we 
derive a generalisation of Noether's theorem for general theories, and show that symmetries
imply a strong restriction on the type of evolution permissable.
With respect to locality, we distinguish violations of causality 
from the creation or destruction of space-like seperated correlations.
We show that violations of causality need not occur
in a relational framework, although one can have  situations where correlations
decay faster than one
might otherwise expect or can be created over spatial distances. This
creation or destruction of correlations cannot be used to signal superluminally,
and thus no violation of causality occurs. 
We prove that theories with information destruction 
can be made time-symmetric, thus impossing no arrow of time.  
}

\author{Jonathan Oppenheim  \\ Department of Applied Mathematics and
Theoretical Physics, University of Cambridge, U.K}
\author{Benni Reznik \\ Department of
Physics
and Astronomy, Beverly and Raymond Sackler Faculty of Exact Sciences,
 Tel-Aviv University, Tel Aviv 69978, Israel.
}
\preprint{}
\begin{document}

\section{Introduction}

All of our current theories of nature are unitary -- an initially pure quantum state 
evolves into another pure state.  
Evolution is completely predictable and reversible.  But it doesn't have to be this 
way -- one can construct 
theories in which pure states evolve irreversibly into mixed states -- information 
is destroyed, and predictability
breaks down.  If we follow
the dictum: ``That which is not forbidden is required'', then we ought to seriously 
consider such non-unitary 
theories unless they
are ruled out by other considerations.  Indeed, that such evolution should be 
considered on this ground alone has long been
advocated~\cite{marinov72,marinov74},
but it was the black hole information paradox~\cite{hawking-unpredictability}
that led to serious consideration of non-unitary theories.  Another motivating factor 
may be found in some
interpretations of  
quantum measurement, 
since some interpretations involve non-unitarity in the form of
fundamental or intrinsic forms of decoherence, or spontaneous wave-function collapse. 
Even in some versions
of the many-worlds interpretation, the effective description of one of the branches 
is an effective non-unitary theory.

The motivation coming from black hole physics is most easily seen 
when we create a black hole from an initially pure quantum state and then allow the black hole
to evaporate away into thermal radiation\cite{HawkingOriginal}. Without radical changes to physics as we know it,
 the evaporation
process must be non-unitary.  Either Hawking's calculation holds, and the pure state evolves into a mixed state of thermal radiation, in which case
the evolution must be non-unitary;
or, Hawking's semi-classical calculation breaks down, and information does escape from the black hole through
subtle correlations, in which case the evolution is again non-unitary.
This is because
if evolution takes a state outside of the
light cone (from A to B for example), and our theory is relativistically 
invariant, then there exists a set of hypersurface~\cite{susskind-thorlacius-note-preskill} 
in which the state has evolved from an initial
copy at A, to two copies of the state, one at A and one at B.
Such an evolution cannot be unitary---it violates the no-cloning
theorem\cite{WoottersZ-cloning} which states that no unitary (or even linear)
map can take any unknown state $\psi$ and produce two copies of it: 
$\psi\otimes\tau\rightarrow\psi\otimes\psi$.  
In the case of the black hole, one finds
a space-like hypersurface which is well-away from the singularity,
yet intersects almost all the outgoing Hawking radiation as well as the
infalling matter which formed the black hole inside the apparent horizon.
This hypersurface contains two copies of the state.
Thus if information eventually escapes the black hole,
the no-cloning theorem (and hence, unitarity and linearity), would be violated.
We thus have the amusing situation that if no information escapes from
the black hole, unitarity is violated, yet if information escapes from a black
hole, unitarity once again appears to be violated.

This is not necessarily a paradox -- we are merely forced to conclude that fundamentally,
our laws of evolution might be non-unitary.  The other possibilities, most notably, that 
the casual structure of space-time might get modified even in 
the semi-classical regime, is very radical, although 
there is some evidence for this in the form of the AdS-CFT conjecture~\cite{adscft}.
It might also be that all the information escapes at the very end, when the black hole has shrunken to the Planck
scale, becoming a remnant with arbitrary high entropy 
and all the problems this entails\cite{aharonov-remnants,preskill-infoloss-note} (but c.f. \cite{bhlock}).  

In this article we do not take a strong position on whether the laws of evolution are non-unitary  
-- we simply stress that it is a possibility,
and therefore the consequences ought to be explored much more fully.  It is not enough to reject non-unitary
evolution based on aesthetic grounds.  All of the current candidates for a theory of quantum gravity, such as string-theory or loop quantum gravity, are unitary theories, while a serious discussion of non-unitary theories is strikingly absent.

That a full exploration of fundamentally 
non-unitary evolution has not happened yet, 
can be traced in large part to a result by
Banks, Peskin and Susskind (BPS)\cite{bps}.  There, the authors argued that non-unitary evolution must either be
non-local, or it must violate conservation of momentum and energy.  


This tension between conservation laws and locality has not been overcome, and 
has posed a stumbling block to further exploration of non-unitary theories.  
While not denying that this tension existed, Unruh and Wald\cite{unruh-wald-onbps} argued that if the
non-unitary part of the evolution was at sufficiently high energy, then the violations in momentum
and energy conservation would also be at high energy, and hence, not observable in the lab.   An attempt to implement
such a theory in a natural way has been proposed simultaneous to the current work \cite{poulin-inprep}. Unruh and Wald also noted
that if the evolution laws had a memory, one could restore the conservation laws.

We will not adopt these approaches here, although we feel they have merit and deserve further study.  Rather,
we will argue that locality and conservation laws are not fundamentally incompatible.  
We will begin in Section \ref{sec:bps} by discussing the 
objection of Banks, Peskin and Susskind, and the general form
that non-unitary evolution must have.  We note that their objection also applies to many
interpretations of measurement in quantum theory.
We then show that locality and conservation laws can exist in harmony provided one considers
relational theories.  For example, non-unitary theory which decoheres a particle into its position with respect
to an absolute notion of space will violate momentum conservation, but one which decoheres two 
particle into relative positions will not violate momentum conservation, and can 
still be local.  

For pedagogical reasons, we first discuss momentum conservation and relational degrees of freedom in the 
context of quantum mechanics in Section \ref{sec:qmmomentum} before moving to the full
field theory case in Section \ref{sec:relationalfields}.  However, 
from a conceptual point of view, many
readers will find Sections \ref{sec:bps}-\ref{sec:qmmomentum}
satisfactory for learning the central idea behind our
construction.
Next, we clarify different notions
of locality.  In particular, we note that one ought to distinguish theories which violate
causality, and 
the far less problematic form of non-locality where correlations
might be created or destroyed.  In particular, we find that
the relational theories we are interested in can lead to the creation of 
short-lived correlations between two
spatially separated regions due to indistinguishability of particles in relational theories.  
The existence of these correlations need not violate causality or allow super-luminal
signalling. Additionally, one can have situations where the  decay of correlations is 
slightly faster than what one might expect from our intuitions about 
non-unitary theories.  
Again, there does not appear to be any troubling consequences due to such effects.
Demonstrating that one can have non-unitary theories which satisfy conservation laws
and are causal is done in Section 
\ref{ss:twofields} 
Then in Section \ref{sec:correlations} we show how they may cause correlations to 
behave differently than in Hamiltonian evolution. 
We thus find that non-unitary theories
are not non-local in the traditional sense, but rather have 
a benign and subtle form of non-locality which does not 
lead to violations of causality and appears to be a small effect.  
The theory respects a form of minimal Lorentz covariance, in that
the equation of motion is invariant under the proper orthochronous Lorentz
group.  Lorentz invariant models in the context of non-unitary theories
where first introduced
in \cite{alicki-reldecoherence}.  In Section \ref{sec:lorentz} 
we prove that non-unitary theories
must violate a stronger form of Lorentz invariance, in that the unitary violating
term cannot be a Lorentz four-vector.

At this point, it is worth mentioning that
from a mathematical point of view, one can alway take a non-unitary theory, and 
make it unitary by considering an enlarged Hilbert space\cite{stinespring} where one
adds an environment.  However,
in  Section \ref{sec:causality} it is apparent why the theories described here should be
considered as fundamental.  The environment that needs to be used is one which doesn't
exchange any momentum, energy or other conserved quantities with the system.  It is also
in a highly non-local and artificial state, involving an infinite number of fields, 
and the interaction must also be engineered in a very particular way. So although one
could consider it to be a physical environment which is invisible to ordinary interactions,
this would be an unnatural view, since the environment interacts with the visible world in
a way which is completely unlike any interaction which commonly exist, and incredibly complicated
and engineered.  It is thus far simpler
in such a case, to consider the non-unitary theory to be fundamental.


Having shown that causal non-unitary theories can conserve momentum, 
we next show how to construct theories which
conserve any other conserved quantity.
In contrast to the other conservation laws, conservation of energy needs to be treated differently.  This
is for a number of reasons -- in particular the Hamiltonian is no longer the generator of time translations,
and as a result time-translation symmetry no longer implies as strong a restriction on the dynamics as do
other symmetries.  Nonetheless, we show that time-translation symmetry if applied to the entire universe, does
require conservation of the total energy of the universe.  In Section \ref{sec:energy} we discuss these
restrictions, and construct a local theory which satisfies them.  We also discuss the differences between
Energy conservation/time-translation invariance and other conservation laws, and symmetries.
The latter are treated in Section \ref{sec:general} 

In Subsection \ref{ss:richness}, we discuss the question of how rich are 
the set of observables which can describe relational
quantities locally. This is relevant because these are the set of observables
into which our theories can decohere a system into.  We claim that this set,
which is also the set of observables 
which
commute with all conserved quantities, is a large set, and can thus lead to highly non-trivial
decoherence.  Another way to say this, is that the Hamiltonian, if it describes the
full system and not just a part of it, is highly degenerate and there are thus
many observables which can commute with it.  

One should however ask whether we ought to attempt to satisfy conservation laws.  This is
because Noether's theorem only applies to unitary theories.  For non-unitary theories,
it is not clear whether a symmetry implies a conservation law.
We address this question in Section \ref{sec:noether}, and derive 
a generalization of Noether's theorem for non-unitary theories.    
Surprisingly, one finds that symmetries do in fact, lead to strong constraints 
on the dynamics, although exact conservation laws
may be relaxed in some circumstances.  
These results apply not only to theories which are fundamentally non-unitary, but also to
effectively non-unitary theories which occur when a system interacts with an environment 
but still possesses some
symmetry. We derive the resulting
continuity equation and modified conservation laws for the stress energy tensor.
When coupled to gravity, stronger constraints on the stress-energy tensor may also occur,
and hence one might have stronger constraints on conservation laws.

We then 
turn to the original motivation for considering such theories -- and discuss 
how to couple such theories to
gravity and in particular, black holes. We mention a few routes one might
take in such an endeavor in Section \ref{sec:coupletogravity}.  
We also note that because non-unitary theories can create correlations without
violating causality, they open up possible alternatives to inflation,
for solving 
the horizon problem (the fact that the universe appears homogeneous, even
though distant regions in space are causally disconnected).  
This is discussed briefly in Section \ref{ss:horizon}.

Next, we turn to another motivation, commonly cited for non-unitary theories, namely, 
theories of stochastic collapse.  Such theories attempt to explain
the so-called ``collapse of the wavefunction'' by appealing to a stochastic mechanism.
While we do not necessarily advocate such attempts, we do note  that
the relational theories considered here can solve some of the difficulties which have
plagued stochastic collapse models.  Namely, such models violated conservation laws
and one could not construct a field theoretic version of such models.  We show how to do
so in Section \ref{sec:collapse}.

We conclude in Section \ref{sec:conclusion} with some of the many open questions which remain. 
Throughout the paper, we try to keep the discussion as general as possible, but in Appendix
\ref{sec:models} we discuss a number of particular model theories which may be of interest.
We also show how correlations behave in some of these models in Appendix \ref{ss:decaycormodel}.
In Appendix \ref{sec:retrodiction} we discuss a question which arose in the context of our
discussion on Lorentz invariance, namely, whether such non-unitary models have an arrow of time.
It is generally claimed that this is the case.  However, we show that one can construct a
time-symmetric version of any non-unitary theory.  Given the state of a system at some 
time $t_o$, predicting the future evolution of the state in a non-unitary theory is just
as difficult as retrodicting its past, which is natural in any theory with randomness.

\section{The tension between locality and conservation laws}
\label{sec:bps}

The state of a system in both quantum mechanics and quantum field theory, is described by a density matrix
$\rho$.  Any new evolution laws must evolve the density matrix into another density matrix, and must therefore
be trace preserving and completely positive.  They must also be linear in the density matrix if the meaning
of the density matrix is to be maintained, since a probabilistic mixture of two states ought to evolve into
a probabilistic mixture of each of the evolved states.  The evolution laws are thus completely-positive, trace-preserving
(CPT) maps $\Lambda$.  
We will generally restrict ourselves to theories which do not possess a hidden memory -- the evolution law at time $t$ depends only on the state of the system at time $t$ and not at other times (the evolution is said to be Markovian, or is generated
by a semi-group)
\beq
\label{eq:semi-group}
\Lambda(t_1)\Lambda(t_2)=\Lambda(t_1+t_2)
\eeq
This feature can be motivated by the fact that if our theory is fundamental, there are no additional degrees of freedom
which could be used to store the information about the system's past.  
%
However, one can have theories which are non-Markovian, due to a 
non-locally in time if one looks at very short-time scales.  Discussion of such theories,
as well as other relaxation of the conditions we impose,
is beyond the scope of the present work.  
%

It can be shown that the most general Markovian map for bounded operators\cite{GKS76,Lindblad76} is given by
\beq
\frac{d\rho}{dt}=-i[H,\rho]-\12\sum_{ij}\gamma_{ij}(\gL_j^\dagger \gL_i \rho 
+ \rho \gL_j^\dagger \gL_i -2\gL_i \rho \gL_j^\dagger)
\label{eq:lindblad}
\eeq
with $\gamma_{ij}$ a positive matrix and $H$ is the usual Hamiltonian. Since $\gamma_{ij}$ is positive,
one can always consider it to be diagonal.   This is referred to as the 
Kossakowski-Lindblad master equation, or usually just Lindblad equation\cite{semigroups}.  
The operators $M_i$ are usually referred to as Lindblad operators, and we will adopt this
convention also.  We will sometimes refer to the $M_i$ as observables when they are Hermitian,
because, as we will see, the resulting evolution given by Equation (\ref{eq:lindblad}) is
very similar to the evolution of a system after measurement of the $M_i$.
The term on the right hand side in addition to the Hamiltonian commutator is
sometimes called the dissipater ${\cal D}(\rho)$.
The sum over operators could be replaced by a continuous spectrum
of operators, in which case the sum would be replaced by an integral.  In the bulk of this paper we
will consider the discrete case, since extension to the continuous case is usually straightforward.

Two examples are of particular interest.  
The first is when the $\gL_i$ are 
projectors $P_i=\proj{i}$, and $\gamma_{ij}=\gamma \delta_{ij}$ 
in which case the Lindblad equation 
decoheres the state $\rho$ in the $\ket{i}$ basis, with off-diagonal matrix
elements $\ket{i}\bra{i'}$ decaying exponentially fast to zero at a rate
$\exp{-\gamma t}$.  This is pure decoherence, and is
much like a measurement in the $\ket{i}$ basis (although without collapse into a particular $\ket{i}$).  
A second example, is to take only one operator $\gL$, 
and have it be a self-adjoint operator (i.e. an observable).  This will also cause decoherence in the eigenbasis
of the operator $\gL$, with a decay exponent for matrix elements 
$\ket{j}\bra{k}$ given by $(m_j-m_k)^2$, with $m_j$ being the 
eigenvalue of $\gL$ in eigenstate $\ket{j}$.  In the case of Hermitian Lindblad
operators $\gL_i$ after diagonalization of $\gamma_{ij}$, 
we can rewrite the Lindblad equation as
\beq
\frac{d\rho}{dt}=-i[H,\rho]-\12\sum_{i}\gamma_{i}[\gL_i,[\gL_i, \rho] 
\label{eq:lindbladhermitian}
\eeq

Mathematically, one can think of such maps as being the result of a unitary
evolution acting on the system and a hidden environment.  
However, we shall see that for the theories considered here, 
it would not make any physical sense to think of it in this way.

Banks, Peskin and Susskind (BPS) rediscovered 
the master equation (\ref{eq:lindblad}), and argued that its evolution,
if local, will lead to a violation of both
energy and momentum conservation. 
We will not reproduce their result here, but rather give intuitive arguments which contains the
core ideas.  
A particular example of this Lindblad evolution would be the proposal of Hawking \cite{hawking-unpredictability}
which BPS recast as
\beq
\frac{d\rho}{dt}=-i[H,\rho]-\frac{a}{2m_p^4} \int dx [F^{\mu\nu}F_{\mu\nu}(x), 
[F^{\tau\sigma}F_{\tau\sigma}(x) ,\rho ] ]
\label{eq:hawkingevolution}
\eeq
where we have simply taken the Lindblad operators $F^{\mu\nu}(x)$ to be 
the electromagnetic tensor.
Such Lindblad operators are local, and the evolution law 
leads to massive production of energy and momentum, 
even if the initial state is the vacuum.  It is not difficult to see why.
The operators $F_{\mu\nu}(x)$ do not commute with the total momentum $P$ of the field.  
The evolution corresponds to the system interacting
with a local environment, and each interaction gives the particles a kick of momentum.  

We can see this clearly in the Heisenberg
representation -- the evolution of operators $G$ which are conserved by the usual Hamiltonian evolution, will instead
evolve under the
master equation as
\beq
\frac{dG}{dt}=-\12\sum_{ij}\gamma_{ij}(\gL_i^\dagger \gL_j G 
+ G \gL_i^\dagger \gL_j -2\gL_i^\dagger G \gL_j)
\label{eq:lindbladg}
\eeq
If we demand that $G$ is exactly conserved, 
then the $\gL_i$ ought to commute with it.  However, when 
$G$ is the momentum $P$ of the field, only highly non-local operators $\gL_i$ can commute with $P$.  Thus,
BPS argued that non-local operators $\gL_i$ will result in non-local dynamics.

There is a connection here with quantum measurements.  
Closely related to the evolution of Equation (\ref{eq:hawkingevolution}) --
not in the context of field theory but within quantum mechanics -- is to
consider Lindblad operators which are projectors onto the position $x$, $P_x=\proj{x}$ leading to evolution
\beq
\frac{d\rho}{dt}=-i[H,\rho]-\gamma \int dx P_x \rho P_x
\label{eq:lindbladx}
\eeq
Again, one can think of the particle as interacting with a local environment at each point $x$, and the momentum of
the particle will be disturbed.
Such a master equation leads to decoherence which one can think of as very similar to a measurement. 
The evolution given by Eq. (\ref{eq:lindbladx}) decoheres a particle into position eigenstates.  Although
there is no collapse, it is as if the position of the particle is being measured.  
A measurement of position disturbs the momentum of the particle, since 
\beq
[X,P]\neq 0 \s .
\eeq  
In much the same way as Hawking's evolution Equation (\ref{eq:hawkingevolution}) interacts locally and thus disturbs 
the momentum, so a position measuring device interacts locally with particles 
and generally disturbs their momentum (c.f. \cite{aharonov-bohm-energy-time}).

There is thus something disturbing about the conclusions of BPS.  Although they are considering evolution
laws, the exact same considerations would seem to apply to measurements. Depending on one's interpretation
of measurement, one may be forced to conclude that momentum is not conserved.
However, we locally measure the position of particles all the time, and we don't believe we are violating momentum
conservation.  But if $P$ is changed when we measure $X$, how is it that we can measure the position of a particle without
violating momentum conservation?  

The answer is that we don't actually measure $X$ of the system with respect to
some absolute reference frame, but rather we measure $X_S$ of the system relative to the position of
some measuring apparatus $X_R$ which acts as a reference system\cite{aharonov-susskind}.  We can thus measure the relative position $X_S-X_R$, 
without disturbing the total momentum $P_S+P_R$ of the system plus reference, since
\beq
[X_S-X_R,P_S+P_R]=0
\eeq
This suggests that we should consider Lindblad operators which are relational.  We will now see that relational
Lindblad operators can allow one to have local interactions which conserve momentum. 
Before moving to the modified evolution laws in the case of 
quantum field theory, it will be
instructive to first look at the problem in the context of 
ordinary quantum mechanics.  After using this as a toy model we will proceed to evolution in the context of 
quantum field theory, which will allow us to examine locality issues in greater detail.

\section{Relational evolution within quantum mechanics: local dynamics with momentum conservation}
\label{sec:qmmomentum}

As an example, we wish to produce a theory which decoheres particles into
position eigenstates.  Normally, this would be given by Lindblad operators of the form
$P_x=\proj{x}$ resulting in Equation (\ref{eq:lindbladx}) above.  Since this violates momentum conservation let us add a second particle 
(a reference particle) and consider instead the Lindblad operator
\beq
\label{eq:coincidence}
Q=\int dx \proj{x_1}\otimes\proj{x_2}
\eeq
and evolution of the form of Equation (\ref{eq:lindblad}), but with non-relational Lindblad operators $\gL_i$ replaced by
the relational operator $Q$.  In analogy with operations sometimes 
performed in the context of quantum information theory\cite{BBPSSW1996}, we will say that $Q$ is created from $\proj{x}$ by {\it group averaging}.
In quantum information theory, this is sometimes called {\it twirling}.  We will
henceforth use $Q_i$ to denote a Lindblad operator which is relational, while
general Lindblad operators will be denoted by $\gL_i$.  We will also have
need for local Lindblad operators in field theory.  These we will denote 
by $L_i(x)$. 

Using $Q$ in place of $\gL$ leads to evolution of the form
\beq
\frac{d\rho}{dt}=-i[H,\rho]-\12\gamma(Q \rho 
+ \rho Q -2Q \rho Q)
\label{eq:lindbladqmm}
\eeq
Q is the relevant operator for coincidence measurements, i.e. it corresponds to the observable that two particles
are at the same position.  One can think of the first particle as being the detector, and the second particle as
being the system (or visa versa).  $Q$ then measures whether the detector ``clicks''.  By using this relational operator 
$Q$ as a Lindblad operator in Equation \ref{eq:lindblad} we get a theory which only decoheres two particles when
they are in the same place.  At all other times, the operator $Q$ does nothing, and one has ordinary Hamiltonian 
evolution.  This evolution is thus local.  On the other hand, total momentum 
$P_{tot}=P_1+P_2$ is conserved.  This is easy to see by
rewriting $Q=\int dx e^{-iP_{tot}x}\proj{x_o}\otimes\proj{x_o}e^{iP_{tot}x}$ which clearly commutes with the
translation operator $e^{-iP_{tot}y}$ ($Q$ is actually invariant under it's action due
to the bounds of the integral being from minus infinity to positive infinity), 
and hence $[Q,P_{tot}]=0$.

Let us now modify the Hawking model of Equation (\ref{eq:hawkingevolution}) 
to ensure momentum conservation.  To do so in the
context of a quantum mechanical reference system, we add 
a non-relativistic object.  Rather than
taking $F^{\mu\nu}(x)$ as the
Lindblad operators, we take the relational operator
\beq
Q=\int dx F^{\mu\nu}F_{\mu\nu}(x)\otimes\proj{x}
\label{eq:fieldtoparticle}
\eeq
and evolution of the form 
  \beq
\frac{d\rho}{dt}=-i[H,\rho]-\frac{a}{2m_p^4} [Q^{\mu\nu}Q_{\mu\nu}, [Q^{\tau\sigma}Q_{\tau\sigma} ,\rho ] ]
\label{eq:hawkingevolutionmod}
\eeq
Because of the integration over all space, it is again easy to see that the $Q^{\mu\nu}$ commute with the total momentum
of the field and reference system.  It is also local --
the difference between this evolution and the initial one is that decoherence only occurs at the location 
where the reference system is, and is zero everywhere else. If the reference system is at a particular location such that the total state of the system is initially 
\beq
\rho=\rho_{field}\ot \proj{x_0}
\eeq
and the particle is heavy enough that the wavefuntion doesn't spread fast,
then
Equation \ref{eq:hawkingevolutionmod} gives
\beq
\frac{d\rho_{field}}{dt}=-i[H,\rho]-\frac{a}{2m_p^4} 
 [F^{\mu\nu}F_{\mu\nu}(x_0), [F^{\tau\sigma}F_{\tau\sigma}(x_0) ,\rho_{field} ] ]
\eeq
which looks exactly like Equation(\ref{eq:hawkingevolution}) for the field degrees of freedom.

Note that the above evolution law may not be too far from what we are aiming for.  It produces decoherence at the
place where our reference object is. Although a black-hole is not a particle, one might imagine that it
acts very much like a quantum particle for the purposes of 
these considerations and that something like the above evolution law might be what we
are seeking with $\proj{x}$ being approximately the position of the singularity.

It should also be noted, that although total momentum is conserved, the momentum
of each system need not be.  For the relational Lindblad operator of
Equation (\ref{eq:coincidence}), when both particles are coincident, and
hence undergoing decoherence, they will get a momentum kick in opposite 
directions.  This can be lessened either by making $\gamma$ sufficiently
small, or  
by smearing the projection operators, for example, instead of
$\gL_x=\proj{x}\otimes\proj{x}$, we can take it to be a smeared projection
\beq
\gL_x=\int dy g(x-y) \proj{y}\otimes\proj{x}
\eeq
with $g(x-y)$ either Gaussian $\exp{[-\Delta(x-y)^2]}$or some other peaked function around $x$.  This means that there will be violations of causality
on the scale of the width of $g(x-y)$ (this is the same degree of causality
violations which appear in the original spontaneous collapse theories which we
discuss in Section \ref{sec:collapse}).  The momentum exchange between individual particles 
can be lessened more easily in  a field theory,
a point we will return to in Section \ref{sec:relationalfields}.  However, on
a fundamental level, one should not be concerned about momentum exchange (it of course
happens in unitary theories as well), except to the extent which it may violate experimental
constraints.

Within quantum mechanics,
we can perform group averaging on any local operators, either smeared or not, 
the idea is just to build in 
some redundancy with extra particles so one has non-trivial Lindblad
operators.  We now move to a discussion
of similar dynamical models as the one given above but in the context of quantum field theory.  Once
we have shown that one can conserve momentum while still preserving locality in field theories, we will move to the 
more complicated case of energy conservation.

\section{Relational evolution for fields: momentum conservation}
\label{sec:relationalfields}


Consider a {\it raw} local Lindblad operators $L_i(x)$ 
equivalent to the local operators
implied by Hawking's requirements.  

Because these are local
in absolute space $x$, they
cause large violations of energy and momentum conservation.  To conserve momentum, we proceed as in the single 
particle case -- we introduce a second field, and consider relational Lindblad operators 
\beq
\Q{ij}(t)=\int d{\bar x} \lone{{\bar x},t}{i}\ot\ltwo{{\bar x},t}{j}
\label{eq:twofieldmomentumconserving}
\eeq   
and evolution of the form
\beq
\frac{d\rho}{dt}=-i[H,\rho]-\12\sum_{ij}\gamma_{ij}(Q_{ij}^\dagger Q_{ij} \rho 
+ \rho Q_{ij}^\dagger Q_{ij} -2Q_{ij} \rho Q_{ij}^\dagger)
\label{eq:lindbladq}
\eeq
The operator $\Q{ij}(t)$ clearly commutes with the total 
momentum just as the $Q$ of Equation \ref{eq:coincidence} did.  We will see in the next section 
that the evolution is local.  We will take $\gamma_{ij}$ to be diagonal (we could have just as
easily considered the non-diagonal case).  

It is also not hard to see that in the non-relativistic limit, 
evolution of the form of Equation (\ref{eq:lindbladq}) can yield the evolution of 
Equation (\ref{eq:lindbladqmm}) described in the context of ordinary quantum mechanics.  We show this in Appendix
\ref{sec:models}.

We can consider more general forms of dynamics than that which would be given by 
Equation (\ref{eq:twofieldmomentumconserving}).  For example, one could use only one field
and make it ``self-referential'', as is done in one of the models
in Appendix \ref{sec:models}.  
Or we can consider any local operator $L_i(x)$ which can act on as
many fields as we wish, or be a product operator or not, and then 
perform group-averaging it to make
a relational Lindblad operator $Q_i(t)$
\beq
\Q{i}(t)=\int d{\bar x} L_i({\bar x},t)
\label{eq:genmomentumconservingq}
\eeq   
and evolution of the form of Equation (\ref{eq:lindbladq})
Again, this clearly conserves momentum, for the same reason as the two previous examples (i.e. $\Q{i}(t)$
is invariant under the action of $e^{iPx}$).  

Note that although the total momentum is conserved, there will be momentum exchange
between the two fields under such models.  While this is not an issue from a fundamental
perspective, one would like to have some control over it.  We can do this by making
the Lindblad operators less projective.  One procedure is as follows: 
for a local Lindblad operator $L(x)$, let
us first smear it over a small test function in space which has support on a very small point
in space.  This is the usual smearing over a tiny region which just keeps the theory finite, and we will generally assume in this paper
that all field operators are suitable smeared.  We then take the spectral decomposition
of the smeared $L(x)$, and we call the projectors $P_l(x)$ onto the eigenvalues $l$ of $L(x)$.  
We can now consider another set of local
Lindblad 
operators
\beq
{\Pi}_i(x)=\int dl g_i(l) P_l(x)
\eeq
with $g_i(l)$ a highly peaked function around some value of $l$.  This new set can be as degenerate as
we wish (for example, if  $g_i(l)$ is a constant function, then $\Pi_i(x)$
will just be proportional to the identity.   It will thus not disturb the individual systems
at all.  By choosing  $g_i(l)$ appropriately, we can control the  momentum
exchange by making the operator ${\Pi}_i(x)$ more or less degenerate via tuning 
$g_i(l)$ to be more or less peaked.  
We can then use $\Pi_i(x)$ or some function of it, in our group-averaged 
Lindblad operators.  E.g. for
two fields
\beq
Q_{ij}=\int dx \Pi_i^{(1)}(x)\Pi_i^{(2)}(x)
\eeq

We now investigate the theory's locality properties.  We will find
that the theory respects causality if the $Q_i$ satisfy some simple conditions.

\section{Locality of the field equations}
\label{sec:causality}

Although BPS did not specify what they meant by locality, it will prove useful to consider two different notions.  
The most important one is that the equations of motion ought to be casual, i.e. they should not allow
super-luminal signalling.  In this sense, we will see that 
the evolution of Equations (\ref{eq:lindbladq}-\ref{eq:genmomentumconservingq}) is local as long
as it satisfies certain conditions -- for example, 
that the Lindblad operators $L_i({\bar x},t)$ are local, and that they are either
Hermitian or come in pairs with their Hermitian conjugate.  In Section \ref{sec:lorentz}
we will see that not only does the theory satisfies causality, but also a 
form of Lorentz invariance.

However, as noted in the Introduction, the causal evolution of the fields is not the only aspect of locality.  
There is also the question of how correlations evolve. 
As in \cite{bps} we need to consider how spatially separated operators behave
under non-unitary evolution.   We will see that correlations can be created between
space-like separated regions, however since these
correlations cannot be used to signal between two different regions at 
speeds faster than light, there
creation doesn't violate causality and doesn't overturn any cherished physical principles.

The effects of the creation of distant correlations 
are also experimentally benign -- 
the only consequence is that in special circumstances, 
correlations decay slightly faster than one might expect, or can be created. Some
constraints from causality on how correlations can change in the context of EPR experiments
has been explored in \cite{grunhaus-correlations}.

We will first in Subsection \ref{ss:twofields} show that causality can be preserved by 
constructing a model which is derived from a fully
Lorentz invariant field theory by tracing out some degrees of freedom.  Since the original theory is causal (and even
Lorentz invariant),
and the Lindblad equation is derived merely by ignoring some of the fields, the Lindblad equation clearly
respects causality; ignoring degrees of freedom cannot result in super-luminal signalling.  The construction
works for any Lindblad equation whose operators $Q_i$ are integrals of local field operators. 
The field theory and associated environment are very artificial (and apparently, necessarily so), 
and thus should not be thought of as an actual fundamental theory from which the Lindblad equation is
merely an effective description.

Next, we explore, In Subsection \ref{sec:correlations} how correlations evolve.  
The issue of Lorentz invariance is taken up in Section \ref{sec:lorentz}.

\subsection{Causal and relational evolution}
\label{ss:twofields}

Causality requires that for local field operators
\beq
[A(\bar{x},t),B(\bar{x}',t')]=0
\label{eq:causality}
\eeq
when $(x-y)^2 < 0$, and the evolution of $A(\bar{x},t)$ and $B(\bar{x}',t')$ are given by the Lindblad
equation.
We shall now show that relational evolution of the form of 
(\ref{eq:lindbladq}-\ref{eq:genmomentumconservingq})  does not necessarily imply a violation of 
relativistic causality.   More precisely, we show that this evolution is causal
to at least second order in the Born approximation. 
To this end, we consider an explicit construction of a Lindblad model starting from a full model of a system and a (fictitious)  environment which together are fully relativistic and therefore causal.

Consider then a number of fields $\phi_i(x)$ which are coupled to some environment of relativistic fields $\psi_\alpha(x)$
via some Lorentz invariant interaction. 
The full reduced dynamics, obtained  by tracing out the environment must therefor be causal.

Our construction takes an interaction between the fields given by
\beq
H_I =  \int L(x)\Lambda(\psi_\alpha(x)) d\bar{x},
\label{eq:fictionalinteraction}
\eeq
where for simplicity we only take one local operator $L(x)$ which acts on the $\phi_i$,
and with $\Lambda(\psi_\alpha)$ denoting some local operator built of the environment fields $\{\psi_\alpha\}$.  One can
consider many different $L_i$ by just including more interaction terms.

The equation of motion for the full density matrix in the interaction picture is given by
\beq
\frac{d\rho(t)}{dt}=-i[H_I(t),\rho(t)]
\label{eq:interactionpicture}
\eeq
with an integral form of 
\beq
\rho(t)=\rho(0)-i\int_0^t ds [H_I(s),\rho(s)]
\eeq
which can be inserted into Equation (\ref{eq:interactionpicture}).  Tracing out the environment
degrees of freedom in which the $\psi_\alpha(x)$ live, and associated with
the density functional $\rho_E$ yields the equation of motion for $\rho_S$, the density
functional of the fields  $\phi_i(x)$
\beq
\frac{d\rho_S(t)}{dt}=-\int_0^t ds \tr_E [H_I(t),[H_I(s),\rho(s)]]
\label{eq:interactionpicture2}
\eeq
where we have assumed that 
\beq
\tr_E[H_I(t),\rho(0)]=0
\eeq
We will see that we can make the interaction weak, and in such a way that
\beq
\tr_E H_I \rho(t) \approx \tr_E H_I \left( \rho_S(t)\otimes \rho_E\right)
\label{eq:borncond}
\eeq
Note that this condition is often written as
\beq
\rho(t)\approx\rho_S(t)\otimes\rho_E
\eeq
but strictly speaking there will be correlations between the system and environment; it
is just that the interaction term is not sensitive to them.  An example of this is an atom coupled
to a field where the correlations between the atom and field (in the form of emitted photons)
are carried off to infinity and don't interact again with the atom.

Inserting the form of the interaction Hamiltonian, Equation 
(\ref{eq:fictionalinteraction}) 
and the approximation Equation 
(\ref{eq:borncond}) into
the equation of motion for the reduced density matrix, 
Equation (\ref{eq:interactionpicture2}) then
becomes
\beq
{d\rho_S(t)\over dt} = - \int_0^\t ds \int d\bar{x} d\bar{x}'  \langle \Lambda(\bar{x},t)\Lambda(\bar{x}',s)\rangle
[L(\bar{x},t), [L(\bar{x}',s),\rho_S(s)]].
\label{eq:afterborn}
\eeq
This is in fact the second order Born approximation and
accounts only for the lowest order corrections
to unitarity. Interestingly however, in the particular case that the state of the 
system and environment is Gaussian  the above form can be shown to be exact\cite{cummulant}.
We then take $t$ to infinity to remove explicit dependence on the initial time.  This approximation
is valid as long as Equation (\ref{eq:borncond}) holds approximately (i.e. the system is Markovian enough).
This also will be shown to be justified by suitable choice of the interaction and environment.
Rewriting the above integral then gives
\beq
{d\rho_S(t)\over dt} = - \int_0^\infty ds \int d\bar{x} d\bar{x}'  
\langle \Lambda(\bar{x},t)\Lambda(\bar{x}',t-s)\rangle
[L(\bar{x},t), [L(\bar{x}',t-s),\rho_S(s)]].
\label{eq:afterborn2}
\eeq

We must now depart from the standard microscopic derivation of the Lindblad equation.  
It is easy to see that consistency with momentum conservation 
imposes requirements on the environment field correlations.
If we demand translation invariance (i.e. Lindblad operators of the form
Equation (\ref{eq:genmomentumconservingq}), then the environment needs to have constant 
spatial correlations.
We also need fast decay of temporal correlations to reduce Equation (\ref{eq:afterborn2}) 
to the Lindblad form of Equation (\ref{eq:lindbladhermitian}).
Namely, we require  
\beq
\langle \Lambda(x,t)\Lambda(x',t-s) \rangle = \gamma \delta(t-s) \s.
\label{eq:envirocorrelation}
\eeq
Equation (\ref{eq:afterborn2}) then yields the traditional form of a Lindblad equation with Hermitian
Lindblad operators
\beq
\frac{d\rho}{dt}=-\12\gamma[Q,[Q, \rho] 
\label{eq:lindbladqfromtwofield}
\eeq
with the momentum conserving Lindblad operator being identified as
\beq
Q= \int L (\bar{x}) d\bar{x} \s .
\eeq

Clearly any choice of a local system operator $L(x)$ will also conserve momentum.

The question is then can a Lorentz invariant theory manifest 
correlations such as in 
Equation (\ref{eq:envirocorrelation})?
Under familiar situations, such as the vacuum or thermal states,  the field correlations decay both in space and time, thus the environment here cannot be taken as a familiar field state.
However, we can design an interaction and environment with such properties by considering 
an environment with an infinite number of fields  $\psi_\alpha$ at hand.  We can then tailor an interaction in 
such a way that at each instant in time, the field interacts effectively with only one environment field:
\beq
\Lambda(x,t) \sim \int d\alpha \delta(t-\alpha) \Lambda_{\alpha}(x,t).
\eeq
with $\Lambda_{\alpha}(x,t)$ a function of $\psi_\alpha$ only.
As a result, for $t\neq s$, 
the temporal correlations disappear
and we have $\langle \Lambda(x,t)\Lambda(x',s) \rangle = \langle \Lambda(x,t)\rangle\langle\Lambda(x',s) \rangle $.
We can always choose  the field and interaction terms such that 
$\langle \Lambda_t(x,t)\rangle=0$ while $\langle \Lambda_t(x,t)^2\rangle\neq 0$ so that 
Equation (\ref{eq:envirocorrelation}) is satisfied.

We further need to ensure that there are constant spatial correlations in the environment.  We can
choose the environment state to be in a superposition of states which are constant over space
\beq
|\psi_\alpha\ra = \int d\psi g_\alpha(\psi_\alpha) \Pi_x |\psi_\alpha\rangle_x,
\eeq  
where $|\psi_\alpha\ra_x$ denotes the field state at the location x.  This is sufficient to enable
Equation (\ref{eq:envirocorrelation}) to be satisfied.   For example, one can take $\Lambda_\alpha(\bar{x},t)$ 
to be $\psi_\alpha(x,t)$ and $g_\alpha(\psi_\alpha)$ chosen symmetric around $\psi_\alpha=0$, so that 
$\langle \Lambda(x,t)\rangle=0$ while $\langle \Lambda_t(x,t)^2\rangle$ is a constant and independent of $x$.

The free environment Hamiltonian is irrelevant for us and can be set to zero for simplicity.  
Since the system interacts with different degrees of
freedom at each time, the temporal correlations vanish and the Markovian approximations we have made can be justified.
Note however that the Lindblad equation only arises up to second order in perturbation theory.  It is
possible that we might not get an exact Lindblad equation from a causal model. 
However, our model very closely approximates Lindblad evolution, thus the relational Lindblad
evolution is causal, or at least closely approximated by a theory which is. 
Up to this degree of approximation, we see that our relational theory of Equations (\ref{eq:lindbladq}-\ref{eq:genmomentumconservingq}) can be derived from a fully causal and Lorentz invariant theory, and is thus causal.

Note that the construction above works for Hermitian operators $Q_i$, but one can construct an interaction Hamiltonian
which produces non-Hermitian $Q_i$, as long as the $Q_i$ come in pairs with their Hermitian
conjugate. This is achieved via the interaction
\beq
H_I =  \int \left( L(x)\Lambda^\dagger(\psi_i(x))+ L^\dagger(x)\Lambda(\psi_i(x)) \right) d\bar{x} 
\label{eq:fictionalinteractioncomplex}
\eeq
with $\Lambda(\psi_i(x))$ a bosonic operator. 
We then require, instead of Equation (\ref{eq:envirocorrelation}), that
\beqa
\langle \Lambda(x,t)\Lambda(x',t-s) \rangle 
& = & 
\langle \Lambda^\dagger(x,t)\Lambda^\dagger(x',t-s) \rangle
=
0 
\nonumber\\ 
\langle \Lambda(x,t)\Lambda^\dagger(x',t-s) \rangle 
&=&
\langle \Lambda^\dagger(x,t)\Lambda(x',t-s) \rangle 
=
\gamma \delta(t-s) \s .
\label{eq:envirocorrelationcomplex}
\eeqa
This can be done for example, if $\Lambda(x,t)$ is a complex field operator, and then
charge conservation will require that terms like $\langle \Lambda(x,t)\Lambda(x',t-s) \rangle$
vanish.
We can then use a  similar procedure to get the Lindblad equation
\beq
\frac{d\rho}{dt}=-\12\gamma(Q^\dagger Q \rho 
+ \rho Q^\dagger Q -2Q \rho Q^\dagger
+ Q Q^\dagger  \rho 
+ \rho Q Q^\dagger -2Q^\dagger \rho Q)
\label{eq:lindbladqfromtwofieldnonherm}
\eeq
where again 
\beq
Q=\int dx L(x)\s .
\eeq
Equation (\ref{eq:lindbladqfromtwofieldnonherm})
is of the form of a Lindblad equation, with two Lindblad operators
$Q$ and $Q^\dagger$.  Thus, using a relativistic
environment to derive causality,
we get either Hermitian Lindblad operators, as in Equation 
(\ref{eq:lindbladqfromtwofield}), or non-Hermitian 
Lindblad operators that come in pairs as in Equation
(\ref{eq:lindbladqfromtwofieldnonherm}).  We will see in 
Section \ref{sec:noether} that these are pairs of annihilation and
creation operators.  In Equation (\ref{eq:lindbladqfromtwofieldnonherm}),
we have only one pair of Lindblad operators, but by having many terms in the 
interaction Hamiltonian, one can have a Lindblad equation of the form
\beq
\frac{d\rho}{dt}=-\12\gamma_i(Q_i^\dagger Q_i \rho 
+ \rho Q^\dagger Q_i -2Q_i \rho Q_i^\dagger
+ Q_i Q_i^\dagger  \rho 
+ \rho Q_i Q_i^\dagger -2Q_i^\dagger \rho Q_i)
\label{eq:lindbladqfromtwofieldnonherm2}
\eeq

If we now try to find a more general condition which will guarantee that a given Lindblad
equation is causal is complicated by the difficulty
in solving the evolution equation for all times.  However, we can more easily explore the question
of whether instantaneous signalling is allowed.  This is just the requirement that the evolution
laws of a local operator should not depend on distant observables.  I.e.
$\frac{dA(x)}{dt}= {\cal L}(A(x))$ needs to be a local operator.  The conditions for this to be true are explored in Appendix \ref{ss:superluminal}. These conditions 
clearly includes the ones used here, but may be more general.

\subsection{Evolution of correlations}
\label{sec:correlations}

Let us now look at the second notion of non-locality, namely, how
correlations may evolve.  
Here we show that the effect of the non-unitary evolution is fairly benign.
Consider two spatially separated  operators $\A{}$ and $\B$, potentially acting on many fields.  To control divergences it should be understood that
we would actually integrate these operators over some test function highly
peaked around $x$ and $y$, but we shall omit this smearing for ease of notation 
We want to check what happens to correlations between these two local observables.
If the evolution does not change the nature of correlations between distant locations, 
then we would
expect that a product observable would evolve as a product.  I.e.
\beq
\frac{d}{dt}[\A{} \B{}]\stackrel{?}{=}\frac{d}{dt}[\A{}]\B{} +
\A{}\frac{d}{dt}[\B{}] 
\label{eq:product}
\eeq
Local Hamiltonian evolution $H=\int dx {\cal H}(x)$, clearly satisfies the above condition. 

Let us now look at how correlations evolve in the non-unitary theory given by the momentum conserving relational model of 
Equations (\ref{eq:lindbladq}-\ref{eq:genmomentumconservingq}),
and taking again $L(x)$ to be a local operator.  We  ignore the Hamiltonian
term since that evolution is purely local.
\beqa
\frac{d}{dt}[\A{}\B{}]
&=&-\frac{1}{2}\sum_{ij}\gamma_{ij}(Q_{j}^\dagger Q_{i}  \A\B
+ \A\B Q_{j}^\dagger Q_{i} -2Q_{j}^\dagger \A\B Q_{j})
\nonumber\\
&=&-\12 \sum_{i,j} \h \int dz dz'(L_{j}^\dagger(z) L_{i}(z')  \A\B
+ \A\B L_{j}^\dagger(z) L_{i}(z') 
\nonumber\\
&&-2L_{j}^\dagger(z) \A\B L_{j}(z'))
\nonumber\\
&=&
\frac{d}{dt}\A{}\B{}
+
\A{}\frac{d}{dt}\B{} + 
V(\A\B)
\eeqa
where 
$V$ is the terms which violates the product evolution form
of Equation (\ref{eq:product}),
and is given by:
\beq
V(\A,\B)=-\sum_{ij}\gamma_{ij} \int dz [L_i(z),\A{}] \int dz[L_j(z),\B{}]
\label{eq:prodviolproj}
\eeq
%
while the local terms are 
\beq
\frac{d}{dt}\A{}
=
-\12 \sum_{i,j} \h  (L_{j}^\dagger(x) L_{i}(x)  \A
+ \A L_{j}^\dagger(x) L_{i}(x) -2L_{j}^\dagger(x) \A L_{j}(x))
\eeq
and similarly for ${\dot \B}$.  
Because $L_i(z)$ is a local operator, $\int dz [L_i(z),\A{}]$
and $\int dz[L_j(z),\B{}]$ in Equation (\ref{eq:prodviolproj}) 
will be a local
operator at $x$ and $y$ respectively.  Note at this point that
$A(x)$ evolves locally and does not depend on any observables or evolution
at $y$.  This is enough to guarantee no superluminal signaling at an instant,
a point which is discussed in greater detail in Appendix \ref{ss:superluminal}.

Putting it all together, we have
\beq
\frac{d}{dt}[\A{} \B{}]=\frac{d}{dt}[\A{}] \B{} +
\A{}\frac{d}{dt}[\B{}] 
-\sum_{ij}\gamma_{ij} \int dz [L_i(z),\A{}] \int dz [L_j(z),\B{}]
\label{eq:notproduct}
\eeq


It is noteworthy, that if we allow distinguishable particles,
such as the models discussed in Section \ref{sec:qmmomentum}, there is
no violation of the product rule (Equation (\ref{eq:product}) )
for the evolution of observables.  
This can be seen as follows. 
Take as the Lindblad operators, those of the kind given by Equation (\ref{eq:fieldtoparticle})
\beq
Q_i=\int \proj{z} L_i(z) dz
\label{eq:lwithdistinguishable}
\eeq
and imagine we have one distinguishable particle.  Then the local observables
which evolve via the disipater are only those which contain operators which act on this particle
such as $\proj{x}A(x)$ and $\proj{y}B(y)$.  However, it makes no sense to speak of the evolution
of $\proj{x}A(x) \proj{y}B(y)$ -- this term is zero unless $x=y$.

%

This has potentially deep consequences 
in a fully relational theory.  In a relational theory of distinguishable
particles, one can describe observables at two points in space by relating
them to two distinguishable particles, one at each of the two points.  
Furthermore, in a non-relational theory, one can, in effect, 
distinguish two identical particles by distinguishing where they are.  
Measurements on an electron on earth commute with measurements made on an
electron on the moon.  However, in a relational theory of indistinguishable
particles, one is bound to have a violation of Equation (\ref{eq:product})
which merely reflects the fact that in a relational theory, we can't use 
absolute space to distinguish two points, and if the field values are identical
at those two points, then there is nothing which can distinguish those two points.  As a result, the evolution laws must act on both those two points without
differentiating between the two of the.

Another interesting consequence of Equation (\ref{eq:notproduct}) can be seen if one takes the limit
that $x\rightarrow y$ and $A=B$.  We then see that under non-unitary evolution,
powers of operators do not necessarily evolve as the power of the time evolved operator.  i.e. 
the relation
\beq
d/dt A^n(x,t) \stackrel{?}{=} n A^{n-1}(x,t)d/dt A(x,t)
\eeq
which holds for unitary evolution, need not hold here.  We do not know of
any dramatic consequences of the failure of this relation, but it might
be worthwhile to explore such effects.

We thus see that although the evolution laws do not violate causality, they do not satisfy Equation
(\ref{eq:product}) I.e. correlations between two regions might be created or destroyed.  
This will be shown more explicitly in Appendix \ref{ss:decaycormodel} when we consider 
particular models.
Such an effect cannot necessarily 
be used to signal superluminally.  Thus, although the
long-range creation of correlations might make us uncomfortable, the fact
that it does not necessarily
lead to violations of causality requires us to take such a possibility seriously.

For the purpose of looking at the evolution of 
long-range correlations, the time evolution
of the mutual information
between two spatial operators is
a good figure of merit. i.e. we are interested in
${\dot I(A(\xv,t):B({\bar y},t))}$ where $I(A:B)=S(A)+S(B)-S(AB)$ and $S(A)$ is the von Neumann entropy of a quantum
state $\rho_A$, $S(A)=-\tr\rho_A\log\rho_A$. Analyzing such a quantity
for a particular model is beyond the scope of the present article, but
would be of interest.  

\section{Lorentz invariance and non-unitary evolution}
\label{sec:lorentz}



Thus far, we have not concerned ourselves with the question of Lorentz invariance, 
although we saw in Section \ref{sec:causality} that causality
was preserved.   This was because we can view non-unitary evolution as coming
from a fully Lorentz-invariant theory, with an environment which we trace out.
The question we now address, is to what extent the new theory is Lorentz invariant -- it may
be that the traced out environment provides a preferred frame which breaks Lorentz invariance.

  
There are various forms of Lorentz invariance one might demand: the first
is that the right hand side of the Lindblad equation ought to transform
like $\partial/\partial t$, so that both sides of the equation transform
in the same way under a Lorentz boost.  This has been referred to as a  
{\it minimal Lorentz invariance } requirement.  It has been claimed 
\cite{srednicki-purity} that even this minimal requirement is impossible to satisfy.  
However, while certainly difficult, 
one can adjust the construction of Lindblad operators and coupling constant
in order to get an equation which is covariant in this first sense.
This is done, for example in 
\cite{poulin-inprep,alicki-reldecoherence,milburn-lorentzinvardec}.

In the context of relational theories, Lorentz invariance arises quite naturally,  at least
in terms of the minimal requirement demanded in \cite{srednicki-purity}.
This will become apparent in Section \ref{sec:energy},
when we see that Lindblad operators
are  Lorentz scalers if they conserve energy and momentum, and are of the form
\beq
{\bar Q}=\int dxdt L(x,t)
\label{eq:txtwirl}
\eeq

This allows one to construct a minimally transforming evolution law
\beq
\frac{d\rho}{d\tau}=-i[H,\rho]-\12\sum_{ij}\gamma_{ij}(\bQ_j^\dagger \bQ_i \rho 
+ \rho \bQ_j^\dagger \bQ_i -2\bQ_i^\dagger \rho \bQ_j)
\label{eq:lindbladcov}
\eeq
where we take $\gamma_{ij}$ to transform as $\frac{d}{d\tau}$.  I.e. $\gamma$ become dynamical, as it is 
in any model where the non-unitarity is induced by tracing out an environment in a unitary
theory.  In a derivation
based on tracing out the environment,
$\gamma$ is related to environmental degrees of freedom.

This ensures that both the left-hand side and right hand
side transform under a Lorentz transformation in the same way.  One can think of it as a redefinition of the 
time derivative with 
\beq
\frac{d}{d\tau}\rightarrow\frac{d}{d\tau}-{\cal D}_0(\cdot)
\label{eq:redifinitiont}
\eeq
with ${\cal D}_0$ just the usual dissipater term taken from Equation (\ref{eq:lindblad}) and transforming in the
same way as $d/d\tau$.  Note that it doesn't matter whether we view this as a redefinition of $d/dt$, or $d/d\tau$.

The requirement of this minimal Lorentz invariance appears to restrict the Lindblad operators to those
which are of the form of Equation (\ref{eq:txtwirl}), and (as we will see in 
Section \ref{sec:energy}, thus imposes energy conservation).  However, other forms
are possible as well, at least if one relaxes the locality requirements.  However, it certainly
appears to give strong weight in favor of energy conservation.


This minimal form of Lorentz invariance, introduced in \cite{alicki-reldecoherence}, uses
the Lindblad equation for the evolution with respect to a time-like vector $a$, i.e.
$\partial_a$, transforms like a Lindblad equation, while, for a space-like vector, $b$, one
has the usual relation
\beq
\partial_b\rho=-i[P_b,\rho]
\eeq
with $P_b$ the momentum in the direction $b$.  For proper orthochronous Lorentz transformations,
time-like vectors $a$ remain time-like, while space-like vectors $b$ remain space-like, and
thus, this difference between spatial and temporal translations is preserved.
There is thus a distinction between time translation (which is not unitary) and spatial translations which are given
by unitary operators, and are in fact constrained to be unitary transformations.  
%
One might wonder whether the term ${\cal D}_0(\cdot)$
in Equation (\ref{eq:redifinitiont}) can transform as a full four-vector.  

We now prove
that it cannot,
provided that 
the usual definitions of energy and momenta are unchanged.
In the Heisenberg representation the equation of motion of a scalar field $\phi(x)$ is then 
\beq
\partial_0 \phi(x)= -i [\phi(x), H] + D_0[\phi]
\eeq
The invariance means that the equation must have the same form
\beq
\partial_{\bar 0} \bar\phi(\bar x)= -i [\bar \phi(\bar x), \bar H] + \bar D_{\bar 0}[\bar \phi(\bar x)]
\eeq
in another reference frame with $\bar x = \Lambda x$.

For contradiction, we now assume that $D_0$ indeed 
transforms as a time component of a four vector $D_\mu$, hence
\beq
\bar D_{\bar 0}[\bar \phi(\bar x)]= \Lambda^0_0 D_0[\phi] + \Lambda^1_0 D_1[\phi]
\eeq
where the boost is for concreteness taken along the "1" direction.

In addition we also have $\bar \phi(\bar x) = \phi(x)$ and $\bar H= 
 \Lambda^0_0 H + \Lambda^1_0 P_1$ and $\bar \partial_{\bar 0} =\Lambda^0_0 \partial_0 + \Lambda^1_0 \partial_1$.
 Substituting, we then get the necessary condition for invariance
 \beq
 \partial_1 \phi =-i[\phi, P_1] + D_1[\phi]
 \eeq
 We observe that the first two terms are equal due to our assumption that $P_1$ is a momentum component 
of a four vector $P_\mu$. I.e. 
$\partial_1 \phi =-i[\phi, P_1]$ is a kinematic identity, which follows
from the definition of  $P^i$ as
\beq
P^i=\int d\bar{x} \pi \frac{\partial \phi(x)}{\partial x_i}
\eeq
and the canonical commutation relations (in contrast
to $[\phi,H]$ which has no such identity associated with it).  
Hence we obtain
 the requirement
  \beq
 D_1[\phi]=0
 \eeq
I.e. $D_0$ cannot be a component of a four vector, and we only have the minimal form
of Lorentz invariance described above.   This ends the proof.

This is not particularly troubling, since
as demonstrated in Section \ref{sec:causality}, it doesn't lead to violations
of causality.  It is as if there is a preferred reference frame which
distinguishes time from space.

We comment that it is possible to choose a nontrivial $D_0$ in such a way that $D_0 [\phi]=0$ vanishes identically  (e.g. $Q=\int \phi dx$) but is non-zero in the general case $D_0[\pi^n]\neq0$. In this case $\phi$  satisfies the ordinary relativistic equation.   Nevertheless non-invariant corrections will then be present in the equations of motion for higher powers of $\phi$
or its conjugate field $\pi$.

For completeness, we note that applying transformations of Equation (\ref{eq:redifinitiont}) 
to the Klein-Gordon 
equation gives for the time-like component
\beq
(\Box + \mu^2)\phi(\x) 
= -{\cal D}^0({\cal D}_0(\phi))-2{\cal D}^0({\partial_0\phi})
\label{eq:lilkg}
\eeq 
and the Dirac equation would become
\beq
i\gamma^\mu\partial_\mu\psi-m\psi=\gamma^0{\cal D}_0(\psi)
\eeq
with the $\gamma^\mu$ being Dirac's gamma matrices rather than
the coupling constants of the Lindblad equation.

\section{Energy conservation}
\label{sec:energy}

As we have noted, energy conservation is in a slightly different category to momentum conservation.  We will show that it can be conserved exactly.  However,
it may also be that it only needs to be approximately conserved.
There are several reasons for this -- the first is that {\it a priori}, 
energy conservation is in a slightly different category to momentum
conservation:
since the generator of time-translations is
now given through the Lindblad equation rather than by the Hamiltonian, Noether's theorem no longer applies -- time-translation 
symmetry does not imply energy conservation.  We will see in Section \ref{sec:noether} that Noether's theorem
is also modified in the case of momentum conservation, but the change is far
less severe.  
The lack of a Noether's theorem is 
both a blessing and a curse -- on the one hand, it implies that we only need to conserve energy approximately
and only to the extent that experiments place constraints on energy conservation.  On the other hand, without
Noether's theorem, it is not immediately clear what prevents us from seeing massive changes in energy.   As was noted in Section \ref{sec:lorentz},
Lorentz invariance may provide some answer to this question (we will see why, further
in this section).  Additionally, constraints
may arise due to locality, or 
due to  coupling the theory to gravity, both of  which demand energy conservation. 
 We will discuss the former in Section \ref{sec:noether} and the latter 
in Section \ref{sec:coupletogravity}. 

For the moment, let us consider constructing an energy conserving Lindblad equation.
Unlike other conserved quantities, this will involve two steps.  The first,
will be to construct group-averaged Lindblad operators as we might do for any other 
conservation law.  In order to ensure locality, we will need a second step, which
is to impose time-translation invariance.  We proceed with the first step: 
we construct a relational Lindblad operator from any other Lindblad operator $\gL_i$ via, 
\beq
{\bar Q_i}=\int dt e^{-iHt} \gL_ie^{iHt} \s .
\label{eq:time-twirl}
\eeq 
and use the notation ${\bar Q_i}$ to henceforth denote time averaging.

Now, also as before, it is useful to have additional degrees of freedom which act as
a reference frame.  In this case, we need a system which behaves as a clock (although
the distinction as to what is a clock needn't be artificial).  
To find the clock observable, we 
find a decomposition of $H$ into $\Pi_\tau$ and $H_o$ such that 
$H = \Pi_\tau + H_o$, with $\Pi_\tau$ conjugate (or perhaps
approximately conjugate) to some observable $\tau$, and such that $\tau$ commutes
with $H_o$.
In that case, $\tau$ 
will correspond to the physical time, as can be verified by it's equation of motion
\beqa
{\dot \tau} &=& -i[H,\tau] \nonumber\\
&\approx& 1
\eeqa
$\tau$
may correspond to some global observable such as the expansion parameter of the universe; it may correspond to
many local degrees of freedom which taken together constitute a clock; it may correspond to a single field. For the
moment we just let it be arbitrary, and potentially some collective degree of freedom. 
We then take
\beq
{\bar Q_i}=\int dt e^{-iHt} Q_i\otimes \proj{0}e^{iHt} \s .
\label{eq:twirlwithclock}
\eeq
where $\ket{0}$ is the initial state $\ket{\tau=0}$ of the clock and $Q_i$ is our
original Lindblad operator (e.g. the momentum conserving relational operators
of the previous section).
It is not hard to see that ${\bar Q_i}=Q_i(\tau)$ i.e. it is equal to the original Lindblad operator at the 
time $\tau$ as measured by the physical clock.  
\beqa
{\bar Q_i}&=&\int dt e^{-iH_0t} Q_ie^{iH_ot}\otimes \proj{t} \nonumber\\
&=&\int dt Q_i(t)\otimes \proj{t}\nonumber\\
&=&Q_i(\tau)
\eeqa
where the last line follows since $Q_i(\tau)\ket{t}=Q_i(t)\ket{t}$.

Any Lindblad operator given by Equation (\ref{eq:time-twirl}) commutes with the total Hamiltonian $H$, since it
is invariant under the action 
\beq
e^{-iH\epsilon}{\bar Q}e^{iH\epsilon}={\bar Q}
\eeq
and therefore conserves energy when used in the Lindblad equation.  
However, such a quantity is rarely local in space. 

This is partly because of the explicit 
$t$ dependence (in the form of a derivative) on the
left-hand side of the original Lindblad equation. Additionally,
although a field operator $\vp(x,t)$
commutes with $\vp(x',t)$ for $x\neq x'$, the symmetric statement is not true: 
$[\vp(x,t),\vp(x,t')]\neq 0$
for $t\neq t'$.  This is crucial, since our derivation in Subsection \ref{sec:causality} of the locality
of the field operators relied crucially on the fact that the Hamiltonian was local, or,
to put it another way, that
\beq
[\int d{\bar z} L_i({\bar z}),\phi({\bar x})]=D_\phi({\bar x})
\eeq
with $D_\phi(x)$ some local function.
We would not obtain a similar expression if we were integrating over time (c.f. 
\cite{temporalordering}).
The ${\bar Q_i}$ are not local in terms of the absolute background space.  

One can think of this problem in another way -- Because the left hand side of the
Lindblad equation is a derivative with respect to absolute time $t$, it is as if the Lindblad 
operator is being measured (or decohering) at some parameter
time $t$ which is typically not equal to $\tau$ . 
As a result, the operator $Q_i(\tau)$ is generally smeared over all space. As an example, consider
the operator $\delta(X(t))\delta(\tau(t))$ -- it is
completely smeared over
all space except at the time t = 0. If $\delta(X(t))\delta(\tau(t))$ appears 
as a Lindblad operator in Equation (\ref{eq:lindblad}) 
then it is as if the operator is being measured 
at an arbitrary time $t$.
 What we really want is that this operator should
be measured at $t=\tau$ not at some other time $t$.  We will now see that it is possible to have the former
situation. 

Namely, we should now impose time-translation invariance.  The entire density matrix of the universe cannot 
depend on the external time $t$ i.e. this parameter time is physically meaningless.  As a result, we must have
\beq
\frac{d\rho}{dt}=0
\label{eq:wdw}
\eeq 
This is similar to the Wheeler-deWitt equation\cite{wheeler-dewitt} for gravitational systems except that the total 
energy of the universe is not constrained to be zero, but rather, some constant.
Since the density matrix is stationary with respect to this external time, the total energy 
of the universe is constant.  As a result, Equation \ref{eq:wdw} is enough to guarantee that
the absolute vacuum state, for example, is stable.

It implies
\beq
-i[H,\rho]-\12\sum_{ij}\gamma_{ij}(\bQ_j^\dagger \bQ_i \rho 
+ \rho \bQ_j^\dagger \bQ_i -2\bQ_i \rho \bQ_j^\dagger)=0
\eeq                                                               

Decomposing $H$ in terms of the clock Hamiltonian, and the rest, gives 
\beq
i[\Pi_\tau,\rho]
=
-i[H_o,\rho]-\12\sum_{ij}\gamma_{ij}(\bQ_j^\dagger \bQ_i \rho 
+ \rho \bQ_j^\dagger \bQ_i -2\bQ_i \rho \bQ_j^\dagger)
\label{eq:lindladtimeless1}
\eeq

If we consider the density matrix as a function of the physical observable 
$\t$, i.e. $\rho(\t, \t' )$, then we can define the total
derivative in terms of the c-number $\tau$
\beq
\frac{d\rho(\t,\t')}{d\tau}\equiv\frac{\partial\rho(\t,\t')}{\partial\t}
+
\frac{\partial\rho(\t,\t')}{\partial\t'}
\eeq
                                               
Since
$i[\Pi_\tau,\rho] =d\rho/d\tau$,
we have
\beq
\frac{d\rho}{d\tau}
=
-i[H_o,\rho]-\12\sum_{ij}\gamma_{ij}(\bQ_j^\dagger \bQ_i \rho 
+ \rho \bQ_j^\dagger \bQ_i -2\bQ_i \rho \bQ_j^\dagger)
\label{eq:lindladtimeless}
\eeq
which is precisely the usual Lindblad equation but with the external, non-physical parameter time t replaced with
the physical observable time $\tau$ .
In terms of the external $t$, the energy is conserved
\beq
\frac{dH}{dt}=0
\eeq                                                                                              
and as long as we take the Lindblad operators to be the time-averaged $\bQ$ of Equation (\ref{eq:twirlwithclock}),
we will also conserve energy with respect to the physical time, ie.
$dH/d\tau = 0$. 

%

In terms of locality, the Lindblad Equation (\ref{eq:lindladtimeless}) is just as local from 
the perspective of physical clock time $\tau$ as the momentum conserving Lindblad equation (\ref{eq:lindbladq})
was in terms of $t$.  In Equation (\ref{eq:lindbladq}), we have $d\rho/dt$ on the left hand side and 
Lindblad operators $Q(t)$ on the right hand side, while in  Equation (\ref{eq:lindladtimeless}) we have
 $d\rho/d\tau$ on the left hand side and the same Lindblad operators, except as functions $Q(\tau)$ of $\tau$.
The locality properties are thus the same.

This proposal that we make is not without it's difficulties, 
however, these difficulties are also present in any unitary theory as well.
 Namely, if we believe that 
quantum mechanics is fully consistent and
applies to the universe as a whole, then Equation (\ref{eq:wdw}) 
will hold, and one will need to find an internal observable to act
as a clock.   Likewise, if we believe that Hawking radiation can be
explained by a unitary theory of quantum gravity, then the
Wheeler deWitt  Equation akin to (\ref{eq:wdw}) holds.  In other words
if we are trying to decide whether our fundamental evolution laws are 
unitary or non-unitary, then both options suffer from the same problems
when it comes to energy conservation and time.

In a unitary theory one encounters various issues of time and energy
conservation, usually studied in the 
context of attempts to apply canonical quantization to
gravity. 
In particular, although it will be possible to find 
an observable $\tau$ which is at least approximately conjugate to some $\Pi_\tau$, 
if it is not perfectly conjugate, then there is some (arguably small) probability that 
the clock will run backwards\cite{unruh-wald-timebackwards}.  There is however, a lot
of freedom in how we decompose $H$ into $H_o$ and $\Pi_\tau$.  Not only can we 
choose it so that $\tau$ acts as a good clock, but we can also choose it so that
the energy exchange between the clock and system is as small as possible.  This provides
an interesting new criteria for decomposing closed systems, for example, in quantum cosmology.
 
Additionally, it appears that the state $\rho$ does not evolve in time $t$.  This is sometimes referred to 
as the ``frozen time'' formalism.  In fact, this is not a good description of what is happening.  Just
as we group-averaged $Q$ in Equation (\ref{eq:twirlwithclock}), we could have 
instead group-averaged the density matrix to get
\beq 
\bar{\rho}=\int dt e^{-iHt} \rho \otimes \proj{0}e^{iHt}
\label{eq:twirledtrho}
\eeq 
so that given any $\rho$ and physical clock observable, we get a state $\bar{\rho}$
which satisfies
\beqa
\frac{d\bar{\rho}}{dt}=0 \s .
\nonumber
\eeqa
There is nothing frozen about $\bar{\rho}$ -- it simply represents a state where we are ignorant
of the absolute value of $t$ i.e. the initial parameter time $t$.  
However, a hypothetical external observer, who knows what the absolute time $t$ is,
would see  the evolution as $e^{-iH(t-t_0)} \rho \otimes \proj{0}e^{iH(t-t_0)}$.  We should thus think of the state
of the universe as actually being in one of the many states in the integrand of Equation (\ref{eq:twirledtrho}),
and evolving in time $t$ as  
$\rho(t)=e^{-iH(t-t_0)} \rho \otimes \proj{0}e^{iH(t-t_0)}$.  It is not frozen at all.
We just don't happen to know the starting time $t_0$
and therefore, which of the $\rho(t)$ in the integrand happens to describe our universe.

In conclusion, we have seen that we can construct a Lindblad equation which conserves
energy and is still local.  In order to do so, we not only needed Lindblad operators
which commuted with the Hamiltonian, but also we needed the absolute time 
$t$ on the left hand side
of  the Lindblad equation to be replaced by a physical time $\tau$.  The theory has to
be completely relational.

\section{General considerations for ensuring conservation laws}
\label{sec:general}

With the exception of time-translation,
the procedure for creating Lindblad operators which respect 
any particular symmetry is straightforward.  One considers
locally gauge invariant operators $L_i(x)$ (which we call the {\it raw operators}).  
For each symmetry that one wishes to respect, one can add an additional field
which acts as a reference frame (although no artificial distinction need be made between the system and reference frame).  We will see that one can also have self-referential models where a field acts as a reference for itself.  
One then considers joint operators on all the fields  $L_\alpha^{(1,2,3...)}(\x)$.  Often, these will be chosen to be of
product form $L_{ijk...}^{(1,2,3...)}(\x)=L_i^{(1)}(\x)L_j^{(2)}(\x)L_k^{(3)}(\x)...$, but this is not necessary.  One then considers the group
generated by all the
generators $G$ of the symmetries, and we average over this group
\beq
Q_\alpha=\int  dg U(g)L^{(1,2,3...)}_{\alpha}U^\dagger(g)
\label{eq:genlinop}
\eeq
with $dg$ taken to be the uniform measure (i.e. Haar) over the group, and $U(g)$
the unitary representation of the group.  Usually $U(g)$ is obtained by
exponentiation of infinitesimal generators $U(g)=\exp{-iGg}$, but this
is not always the case, for example, with diffeomorphisms.

One can also consider what we call {\it coherent group-averaging}.  I.e. operators of
the form
\beq
Q_\alpha=\int  dg dg' U(g)L^{(1,2,3...)}_{\alpha}U^\dagger(g')
\eeq
we discuss these in slightly more detail in Appendix \ref{sec:models}. 

Note that as in the cases considered here, we are often averaging 
over a non-compact group, and thus the operator does not formally converge. For the
purposes considered here, such operators can still be considered
in the Lindblad equation.  A review of this and other technical issues
can be found in the review of \cite{Marolf2002-groupavgreview} in the context of
the quantization of constrained systems.  A review of general group averaging in
the context of work on reference frames and quantum information
can be found in \cite{BRS-refframe-review}.

Note that one might want to consider variations of this such as considering 
operators $L_{ijk...}^{(1,2,3...)}(\x)=P_i^{(1)}(x)P_j^{(2)}(x+l_2)P_k^{(3)}(x+l_2)...$ as long as 
the $l_\nu$ are small enough that the non-locality of the above operator will be unobservable.  
There are several models of this form which we have studied and we discuss a few of 
these in Section \ref{sec:models}.

In general, there appears to be a large set of observables
that we can use as Lindblad operators, and
many local observables which upon integration
give relational observables.  It seems that one can include
all powers of a local observable
along with conjugate observable, e.g. $\phi^n(x)\pi^m(x)$, and then
perform group averaging on them.
It might be useful to study such models in more detail.  

\subsection{Richness of relational observables}
\label{ss:richness}

The question of what physical observables can be described by integrals of local
field operators is important in order to be convinced that one can get 
non-trivial dynamics from the models we have proposed.  This will become
even more important when we discuss stochastic collapse models, since there,
one is attempting to describe all the observables we measure in the laboratory.

If one considers Lindblad operators of the form of Equation (\ref{eq:genlinop}),
it is not immediately clear how rich their structure is i.e. are there enough
of them, that they lead to non-trivial decoherence.   If the only Lindblad operators
where the total energy, total angular momentum, total electric charge etc., then
we would only be able to decohere into observables which were conserved quantities. 
This would not be very interesting.  The question is whether there are many
other observables besides the conserved charges 
which are of the form of Equation (\ref{eq:genlinop})

Since these Lindblad operators 
can also be
understood as relational observables, one can ask whether relational observables
fully describe
the world around us.  Certainly one expects that relational observables do describe
what we observe, since these are the only physically relevant ones -- they are
the ones we measure in the lab. Indeed, when it comes to conservation
laws which are associated with a Gauss's law (such as charge and energy), we
can only measure operators which commute with the conserved 
quantities\cite{strocchi-wightman}. 
Thus, the relational ones (which commute with these conserved quantities),
being the ones we measure in a lab, must
be sufficiently rich.
However, here, we are also  demanding that the observables be local.  Again, local
relational observables would
appear to be sufficiently rich, since in the lab, we are always measuring local quantities.
Given that in the lab we presumably are measuring local observables which must 
commute with conserved quantities, it is reasonable to believe that Lindblad operators
which correspond to those observables are sufficiently rich.

However, it is not completely obvious that such a set could fully describe the types of observations we
make. 
Take for example a particle located
at $x$ and a second one at $y$.  The relational description would be limited to saying
that there are two particles which are a distance $|x-y|$ apart.  This relational
observable, which can be written as
\beq
Q=\int dz\proj{x+z}\otimes\proj{y+z}
\label{eq:relativedistance}
\eeq 
is not local, and thus cannot be used in our Lindblad equation.  
Lindblad operators of the form
\beq
Q_{\alpha\beta}=\int dx A_\alpha(x) B_\beta(x+l)
\label{eq:nonlocallop}
\eeq
might be permissible, but only if $l$ is sufficiently small.  The theory is non-local
on that distance scale.

Now, if one has a pre-existing reference frame, i.e. a field $\phi(x)$ over all space which has
different values at all positions $x$, then certainly it is simple to use such a field
to fully describe all additional fields in relation to it. For example, as a toy theory
one might be able to use a suitably smeared local projector  $P_\phi(x)$ onto values of 
the smeared field $\phi(x)$ and the relational observables
\beq
Q_{i\phi}=\int dx P_\phi(x)\otimes L_i(x)
\eeq
If $\phi$ is such that it uniquely distinguishes $x$, then this acts like a good reference frame
and the observables $Q_{i\phi}$ will be able to describe any observables with respect to this reference frame.  This includes observables of the form of Equation (\ref{eq:relativedistance}).  
However, without a pre-existing reference frame, it is not clear whether all observables can be described in local and relational terms.  Perhaps it is too much to demand that
a relational model be descriptive without a pre-existing reference frame.

Indeed, in a lab setting, we build up our knowledge of the reference frame over a period
of time.  A spark in a spark chamber (a local coincidence measurement) is equivalent to
a position measurement only because the positions of the wires have been predetermined by
making many measurements over a duration of time (e.g. sending photons to bounce off
various parts of the device, and timing their return).  It is thus certainly feasible that
local relational observables, over a period of time, will decohere a system in a non-trivial
manner, and eventually, completely decohere a system into states which, from a physical point
of view, are the states we observe.

In order to pursue such questions further, the Hamiltonian dynamics is likely to be as
important as the Lindblad operators.  For example, if there is an effective potential 
$\sigma(x)\sigma(y) V(x-y)$
between two spins, then the probability of the two spins being parallel or anti-parallel 
 will be a function of the distance between them.  Lindblad operators which
decohere the spins are in a sense decohering different distances, 
particularly if there are many of them.  Such models, and the general question of
richness of local relational observables is one which deserves future study.

Finally, we often try to find an observable which commute with all symmetries,
and whose commutator with a local observable is a local observable -- namely, when
trying to find the Hamiltonian of a system.  The Hamiltonian is an example of
an operator which commutes with conserved quantities, however, renormalizability
and boundedness place more constraints on a Hamiltonian than it does on an observable.  For
a Hamiltonian, one is forced to throw away odd-powered 
terms like $\phi^3$ which are unbounded from below, but such terms can be included in an observable.  Likewise,
for terms which would lead to non-renormalizable theories, it appears that one
can include these in an observable.  However, 
one needs to ensure that 
including such terms as Lindblad operators would 
not result in a non-renormalizable theory, or lead to
other problems.  Understanding this is an interesting question which we feel
deserves further study.

It is worth noting that for a physical system, if one includes perturbations
due to other systems, then the Hamiltonian is usually non-degenerate.
The energy levels, 
unless protected from external perturbations by a symmetry,
will be split by the presence of the external fields.
Thus, decohering a system in the basis of the Hamiltonian is as decohering
as one can get, and the Hamiltonian encodes the full richness of the theory.
Knowing the energy is a complete observation of the system.
However for closed systems, the Hamiltonian is generically expected
to be highly degenerate.  It is this fact which allows us to 
make all sorts of measurements, which still commute 
with the Hamiltonian.

\section{Symmetries and conservation laws}
\label{sec:noether}

Noether's theorem states that for a system which evolves
unitarily, there is an equivalence between
symmetries of the evolution laws, and the 
conservation of some quantity.  When
the evolution is no longer unitary, the connection between
symmetry and conservation laws may break down.  Most symmetries are probably
regarded as being more fundamental than the associated conservation laws --
if the laws of physics were discovered to be different on Earth than on
the moon, we would be unlikely to conclude that the laws of physics changed just because
we moved to some new location -- rather we would more likely 
posit that the presence of the moon or some unseen matter was causing objects
to behave differently.  We therefor ought to demand of our non-unitary theory
that it be invariant under our cherished symmetries (such as spatial and
temporal translations), but we may be less bothered by the fact that it
may violate a conservation law.

However, this presents a potential
problem for non-unitary theories~\cite{gross-unpredictable} since quantities
like momentum appears to be conserved experimentally, and
it would be difficult to explain such a fact without Noether's theorem.
Without a connection between symmetries and conservation laws, there appears
no reason to demand even approximate conservation laws.  The near exact
conservation laws we experience would therefore appear to be an unexplainable
accident in a theory with non-unitary evolution.  We will return to this point
towards the
end of Section \ref{ss:continuity}.

However, although the connection between symmetry and conservation laws
may break down for a general map which takes mixed states to mixed 
states, it need not be the case.  One has to
look at the structure of the map to make such a determination.  We will
therefore re-examine this connection for the general evolution law given
by Equation (\ref{eq:lindblad})
\beqa
{\cal L}(\rho)=\frac{d\rho}{dt}=-i[H,\rho]-\12\sum_{ij}\gamma_{ij}(\gL_j^\dagger \gL_i \rho 
+ \rho \gL_j^\dagger \gL_i -2\gL_i \rho \gL_j^\dagger)
\s . 
\nonumber
\eeqa

We will find that symmetries do still place a constraint on the type of evolution
which is permissible, although the connection with 
conservation laws is modified.
We also make the distinction between symmetries which are associated with
a generator of a unitary transformation (such as spatial translations), and
a symmetry which is not.  This is important, because in the theories
considered here, time-translation is no longer generated by the Hamiltonian,
but is instead given by a non-unitary operator.  

Let us consider symmetry 
under an infinitesimal transformation implemented by $\ug$ with $G$ a Hermitian
operator which generates the transformation.  Symmetry of the evolution laws
implies 
\beq
{\cal L}(\rho)=
\ugd{\cal L}(\ug\rho\ugd)\ug
\label{eq:invariancecondition}
\eeq
I.e. the equations of motion ``commute'' with translation under $G$ -- if we translate
under $G$ and then evolve the system, and then translate back, the evolution 
should not be different.  Stated another way, if we implement a transformation on 
the evolution laws, and also transform to state, then if there is a symmetry,
the transformed state will satisfy the transformed evolution laws.

When applied to the master equation of (\ref{eq:lindblad} in 
the (potentially degenerate) eigenbasis $\ket{g}$ of $G$, invariance under the symmetry yields
\beqa
&&\left(-iH-\12 \sum_i\gamma_i\gL^\dagger_i\gL_i\right)\ket{g}\bra{g'}
+
\ket{g}\bra{g'}\left(iH-\12 \sum_i\gamma_i\gL^\dagger_i\gL_i\right)
+
\sum_i\gamma_i\gL_i\ket{g}\bra{g'} \gL^\dagger_i
\nonumber\\
&=&
e^{-i\epsilon g}\ugd\left(-iH-\12 \sum_i\gamma_i\gL^\dagger_i\gL_i\right)\ket{g}\bra{g'}
+
e^{i\epsilon g'}\ket{g}\bra{g'}\left(iH-\12 \sum_i\gamma_i\gL^\dagger_i \gL_i\right)\ug
\nonumber\\
&&
+
e^{i\epsilon (g'-g)}\sum_i\gamma_i\ugd \gL_i\ket{g}\bra{g'} \gL^\dagger_i\ug \nonumber\\
\label{eq:noethertosatisfy}
\s .
\eeqa
This can only be satisfied if
\beq
[\pm iH -\12 \sum_i\gamma_i\gL^\dagger_i\gL_i,G]=0 
\label{eq:noetherconditionsboth}
\eeq
and additionally
\beq
 e^{iG}\gL_i  = \gL_i e^{iG - i \Delta_i}
\label{eq:noetherconditionsthird}
\eeq
which ensures cancellation of the final term on both sides of 
Equation (\ref{eq:noethertosatisfy})
when used in conjunction with the Hermitian conjugate
of Equation (\ref{eq:noetherconditionsthird}) 
\beq
\gL_i^\dagger e^{-iG}  =  e^{-iG + i \Delta_i} \gL_i^\dagger \s .
\label{eq:noetherconditionsthirdherm}
\eeq
 Here the $\Delta_i$ are real
constants.
Equations (\ref{eq:noetherconditionsboth}-\ref{eq:noetherconditionsthird}) can
be written as the two conditions
\beq
[G,\gL_i]=-\Delta_i \gL_i
\label{eq:noetherconditions1}
\eeq
and
\beq
[G,H]=0
\label{eq:noetherconditions2}
\eeq
with condition (\ref{eq:noetherconditions1}) implying
\beq
[G,\gL_i^\dagger]=\Delta_i \gL_i^\dagger
\eeq
Another way to write the condition of Equation (\ref{eq:noetherconditions1}) is as
\beq
\sum_i\gamma_i\gL^\dagger_i\ket{g}\bra{g'} \gL_i = \sum_i m^*_i(g)m'_i(g) \gamma_i 
\ket{g+\Delta_i}\bra{g'+\Delta_i}
\eeq
Since the effect of the Lindblad equation
is completely characterized by its effect on a complete basis, we can always take  
each $\gL_i$ to individually be proportional
to a raising or lowering operator of $G$ 
\beq
\gL_i^\dagger\ket{g}=m^*_i(g)\ket{g+\Delta_i}
\label{noethercondition4}
\eeq
and hence
we get Equation (\ref{eq:noetherconditions1}). 
This also implies that $\gL^\dagger_i\gL_i$ commutes with the generator of the symmetry.
\beq
[G,\gL^\dagger_i\gL_i]=0
\eeq
thus $\gL^\dagger_i\gL_i$ has the same eigenbasis as the  
number operator. 
Note again that $\ket{g}$ might be highly degenerate, and the values $m^*_i(g)$ can depend
on other degrees of freedom -- we have just dropped them for convenience.

The Lindblad equation in the Heisenberg representation Equation (\ref{eq:lindbladg}), together with conditions
(\ref{eq:noetherconditions1}-\ref{eq:noetherconditions2}) gives
\beqa
\frac{dG}{dt}&=&\sum_i \gamma_i  \gL_i^\dagger [G,\gL_i]
\nonumber\\
&=&\sum_i - \gamma_i  \Delta_i\gL_i^\dagger\gL_i
\label{eq:diffg}
\eeqa
i.e. the rate of change of $G$ is given by an average over the number operator.
We may then write the solution to Equation (\ref{eq:diffg}) as
\beq
G(t)= G(0) - (\sum_i \gamma_i \Delta \gL_i^\dagger\gL_i)t
\eeq


Here, $G(t)$ could grow without bound, although it grows at a linear rate,
rather than exponentially fast. 
From our discussion on causality, we found that the relational Lindblad 
equation was causal when the $M_i$ were Hermitian, or when they came
in pairs with their Hermitian conjugate.  If we take these two cases,
then we have that either
\beq
{\dot G(t)}=0
\label{eq:gratehermitian}
\eeq  
as was the case when the fictitious environment was made up of scalar fields.  Or, 
in the case of complex environment fields, then we have that
for each $M_i$ which is a raising operator by the amount $\Delta_i$, there
is a lowering operator $M^\dagger_i$ by the same amount.  This gave a Lindblad evolution
such as that of Equation (\ref{eq:lindbladqfromtwofieldnonherm}).
This implies that in 
Equation (\ref{eq:diffg}), we must also put in the conjugate terms, which
contribute a $-\Delta_i\gL_i\gL_i^\dagger$ to the sum (since they are lowering operators),
giving
\beq
{\dot G(t)}=\sum_i\gamma_i\Delta_i[M_i^\dagger,M_i]
\label{eq:gratepairs}
\eeq

Comparing Equation (\ref{eq:gratehermitian}) with Equation
(\ref{eq:gratepairs}), one might want to rule out the latter possibility on physical grounds.    
This might be reasonable, given
that non-Hermitian Lindblad operators require a different type of  environment and
interaction Hamiltonian (fictitious or otherwise).  And that such an evolution
would eventually lead to infinite production of the quantity $G$ (although the
time it would take may be very long).  

The only exception to our generalized Noether's theorem 
is for energy which is not generated by the Hamiltonian.
Rather  Equation (\ref{eq:lindblad})
is itself responsible for time-translation, and is clearly time-translation
invariant provided $H$ and  the $\gL_i$ have no explicit time-dependence. 
Thus the only conservation law which requires additional 
explanation is that of energy 
conservation.  We saw however, in Section \ref{sec:energy} that imposing
time translation invariance on the entire state of the universe led to energy conservation. 
 We will also see in the next section that coupling to gravity appears to impose additional constraints.  And finally
 in Section \ref{sec:lorentz} we saw that a weaker form of Lorentz invariance also appears to imply energy conservation, at least for the type of models we considered.

\subsection{The modified continuity equation}
\label{ss:continuity}

We now wish to see how the continuity equation is modified when evolution is non-unitary.  
%
If we have a generator of a symmetry which expands as
$G=\int g(x) dx$ then under ordinary unitary evolution, we would have
\beq
\frac{dg(x)}{dt}=i[H,g(x)] \s .
\label{eq:densityuevolution}
\eeq
Noether's theorem then gives that $dG/dt=0$, which implies that
$
\int \frac{dg(x)}{dt}
$
must be equal to a surface term which vanishes as the surface is taken to
be at infinity.  By Gauss's law, this implies 
\beq
i[H,g(x)]=\nabla_i f^i
\label{eq:Htotaldiv}
\eeq
i.e. the integrand of $ \int dx [H,g(x)]$ can be written as a total divergence.  
For unitary evolution, this implies the usual
continuity equation 
\beq
\partial_\mu f^\mu(x)=0
\eeq
where one defines $f^0(x)\equiv g(x)$.

For non-unitary evolution, Equation (\ref{eq:densityuevolution}) gets modified, and we have
 \beq
\frac{dg(x)}{dt}=i[H,g(x)] -\12\sum_{ij}\gamma_{ij}(\gL_i^\dagger \gL_j g(x) 
+ g(x) \gL_i^\dagger \gL_j -2 \gL_i^\dagger g(x) \gL_j)\s .
\label{eq:densitylevolution}
\eeq
From the symmetry conditions of Equation (\ref{eq:noetherconditions2}) derived in the previous section, we still
have $[H,G]=0$ and thus Equation (\ref{eq:Htotaldiv}) still holds and 
\beq
i[H,g(x)]=\nabla_i f^i
\eeq

For local evolution, we can define
\beq
k(x)=   -\12\sum_{ij}\gamma_{ij}(\gL_i^\dagger \gL_j g(x) 
+ g(x) \gL_i^\dagger \gL_j -2 \gL_i^\dagger g(x) \gL_j).\\
\label{eq:divergenceasdis}
\eeq

From the results of the previous section, we know there are only two cases 
-- either the $\gL_i$ commute with $G$ in which case, once again, $k(x)$ can be written as a total
divergence, and one still has a continuity equation of the standard form.  
This will be the case if $\gL_i$ are Hermitian.
On the other hand, without this requirement, our generalization of Noether's theorem
implies that the $\gL^i$ might also be raising or lowering operators of $G$.  In such a case
the continuity equation gets modified to
\beq
\partial_\mu f^\mu(x)=k(x)
\eeq
and applying Equation (\ref{eq:diffg}) to Equation (\ref{eq:divergenceasdis}) we have that 
\beq
\int dx k(x)=\sum_i\gamma_i\Delta_i gL_i^\dagger\gL_i
\label{eq:divergence}
\eeq

As before, two cases are of special interest due to their locality properties.  The first
is when the $M_i$ are Hermitian, in which case we recover the ordinary continuity equation.
The second case is when $M_i$ comes in pairs with it's Hermitian conjugate, in which case
Equation (\ref{eq:gratepairs}) gives that 
\beq
\int dx k(x)=\sum_i\gamma_i\Delta_i [M^\dagger_i,M_i]
\label{eq:divergenceherm}
\eeq
as is implied by Equation (\ref{eq:gratepairs}).
However, in this case, in addition to the $M_i$ coming in pairs with $M_i^\dagger$, we also
required that the $M_i$ be integrals of local fields in order for the equations of motion to be causal.
\beq
M_i=\int dx L_i(x)
\eeq
Because $L(x)$ is local, the integral of
the commutator $[L_i^\dagger,L(y)]$ is some local function, call it
$N_i(x)\delta(x-y)$.  We then have that 
\beqa
[M^\dagger_i,M_i]&=&\int dxdy [L^\dagger_i(x),L_i(y)]
\nonumber\\
&=& \int dx N_i(x) 
\eeqa
Putting this into Equation (\ref{eq:divergenceherm}),
we thus find that $k(x)$ is a local quantity
\beq
k(x)=\sum_i\gamma_i\Delta_i N_i(x)
\eeq

For the generators of spatial-temporal translations, we have a modified conservation equation
\beq
{T^{\mu\nu}}_{;\nu}=k^\mu
\label{eq:conviol}
\eeq
with 
\beq
{\bar k}(x)= \sum_{i}\gamma_{i} {\bar \Delta}_i N_i(x) 
\eeq
for the $\mu$ which are the spatial components.  I.e.
since our version of Noether's theorem gives no restriction on energy conservation (since $H$ is
no longer the generator of translations), we can impose no restrictions on $k^t$.  However,
the spatial component, $k^\mu$, which we shall denote by ${\bar k}(x)$ must satisfy
Equation (\ref{eq:divergence})
with the possibility that all the $\Delta^\mu_i$ are zero (the case of Hermitian $M_i$, 
in which case
one has ordinary conservation of ${T^{\mu\nu}}_{;\nu}$.  Note that here,
 the ${\bar \Delta_i}$ is now a vector, since it is determined by Equation 
(\ref{eq:noetherconditions1}) which depends on the symmetry generator $G$.
As an example, $N_i$ might be the number operator for 
momentum vector ${\bar p}$.
In this case, the  ${\bar \Delta}_i$ 
which may imply that coupling to gravity 
imposes the condition $k(x)=0$.  We discuss this briefly in Section \ref{sec:coupletogravity}.
It may be that curvature can not couple to matter in
the usual way.  There would also be a modification to the Gauss's law since
energy need not be conserved in an asymptotically flat universe.  Since the gravitational
Gauss's law comes from Einstein's equation, it would necessarily get modified due to
the modification of Equation (\ref{eq:conviol}). 
With regard to the Bianchi identity, a potential resolution may be found in the
model discussed in Section \ref{sec:causality}.  There, one enlarges the Hilbert
space to consider a unitary theory on the original system and an environment.  In such
a model, the violation of stress energy in the non-unitary theory could be understood
as an exchange of energy and momentum with the (perhaps fictitious) environment.  One could
then have full energy-momentum conservation on the extended Hilbert space, and apply
Einstein's equations to the stress-energy due to both the system's matter, and also
the environment's. 
We leave such problems
for further study.  However, it does raise the issue of how we couple our theory to
gravity, a problem which we very briefly address in the following section.

Note that in general one expects that the modification to the continuity equations may be difficult to observe.  
This is because 
the $\gL_i$ are total raising or lowering operators and would not have a big effect locally.  I.e. local functions will be superpositions
over many eigenstates of $G$, and so acting a lowering or raising operator on such a state would not change the state locally, although the
global state will change.

The modification of the continuity equation will lead to a modification of the
Ward-Takahashi identity.  Rather than use the ordinary continuity
equation $\partial_\mu f^\mu(x)=0$ in the identity, we would use
$\partial_\mu f^\mu(x) =k(x)$
 which leads to
\beq
\la\delta_\epsilon{\cal F}\ra = i\epsilon\int\la{\cal F}
\left(\partial_\mu f^\mu(x) -k(x)\right)\ra dx
\eeq
for ${\cal F}$ a functional of the fields and 
$\delta_\epsilon$ the infinitesimal gauge transformation.  
We hope to explore the effect of this modification further.

\section{Coupling to gravity}
\label{sec:coupletogravity}

Despite the fact that one of our original motivations for this study was information destruction
in black-holes, we have thus far only considered non-unitary evolution in the context of flat-space, without
any coupling to gravity.  In all likelihood, coupling such a theory to black-holes will require a quantum theory
of gravity.  Nonetheless, we can construct a few plausible toy models 
in order to understand how such a theory
might couple.  Here me suggest a few different possibilities.  We will not explore them
in any detail, we merely mention some possibilities for future research directions.

{\bf (1) Coupling to the singularity:}
If we want the modification to ordinary physics to only occur at the Planck scale,
or in the presence of black holes, than one is likely to have information destruction
at the singularity of the black hole.
A potential toy model for such a coupling could be something along the lines of
Equations (\ref{eq:fieldtoparticle}-\ref{eq:hawkingevolutionmod}) with $\proj{x}$ projecting
onto the ``position'' of the singularity -- where by position, we understand it to be in
terms of some perturbative expansion around flat space-time.  
\beq
\bQ=\int d\bar{x}dt F^{\mu\nu} F_{\mu\nu}(\bar{x},t)\otimes\proj{\bar{x}}\otimes\proj{t}
\nonumber
\eeq
\beq
\frac{d\rho}{dt}=-i[H,\rho]-\frac{a}{2m_p^4} \int dx [\bQ, [\bQ ,\rho ] ]
\nonumber
\eeq

While this is only a very rough toy model, it has some of the features one might expect
from black holes which destroy information.
This type of coupling is particularly tempting in light of the fact that the quantum black 
hole presumably has a huge number of microscopic degrees of freedom owing to it's large entropy.  
This makes its plausible that it corresponds to some sort of
distinguishable object. As a result, we no longer
have even the mild non-locality discussed in Section \ref{sec:causality},
as discussed following Equation (\ref{eq:lwithdistinguishable}).

{\bf (2) Coupling to the curvature:}

There are many ways one can create Lindblad operators which couple to 
the curavture or the metric in some way, and we mention a few toy models.  For example
\beq
Q=\int dx \sqrt{-g} R(x) L(x)
\eeq
where $L(x)$ is any local field operator and
$R(x)$ is the space time curvature or some other function of the
metric E.g. one could consider vector components of the form
\beq
Q_{\mu\nu}=\int dx \sqrt{-g}R_{\mu\nu} L(x)
\eeq
or 
\beq
Q=\int dx\sqrt{-g} R_{\mu\nu} L^{\mu\nu}(x)
\eeq
or any such combinations, and functions which need not be in product form.  

Such Lindblad operators
would not conserve momentum or energy, but we do not expect conservation
if we were to treat gravity as an external field.  One would have to
take into account the back-reaction on the gravitational field, thus such models
should only be considered as an effective description of the matter degrees
of freedom in a fixed background.  

Note that these models appear to have the required effect of producing decoherence
near black holes, while producing little decoherence in flat-space.  
This is because the curvature terms goes to infinity as we approach the
singularity, and the decoherence terms will dominate, even for very
small coupling constants $\gamma$.

Other relational observables in the context of effective gravity has been discussed
in \cite{giddings2006oeg} and in the context of deSitter space times in
\cite{marolf2008gas}.  Such observables can be used in the relational Lindblad
equation. For example, we can take
\beq
Q_\tau=\int dx \sqrt{-g}f_\tau(R(x))
\eeq
where $f_\tau(R(x))$ is a highly peaked function around $R=\tau$ and can potentially
serve as a clock.

{\bf (3) Black holes as microscopes to high energy physics:}
One can consider a very simple idea, where we don't directly couple gravity to the master equation, but instead
use the fact that a black-hole redshifts energy near the horizon.  
If the dissipater term ${\cal D}(\rho)$ only
acts at high energy, then in ordinary laboratory experiments, one would find 
evolution very close to unitary.
On the other hand, modes which are radiating from a black hole were once at a 
very high energy since they originate from
close to the horizon.  They will therefore have been subject to the  
dissipater term ${\cal D}(\rho)$ causing information
loss.  Whether such a scenario is compatible with the usual picture of 
black-hole evaporation is unclear, but the 
idea is perhaps attractive because of it's simplicity.

{\bf (4) Information Destruction:}
%
We have seen that it is possible to construct relational theories with little constraints
other than locality and conservation laws. This evolution
may be constrained further, by going back to one of our original motivations for
considering non-unitary evolution.  The goal was to use it to destroy information in
the context of black-hole evaporation.  In order to remain consistent with the causal
structure of the black hole space time, no information can escape the black hole until
such time as the black-hole is of Planck size, at which point the causal structure
may break down.  This means that virtually all the information which goes into
the black hole must be destroyed.  We therefore demand of our evolution, that it be
strong enough to completely destroy information.

What we mean by this is that for $\rho_{out,in}$ the density functional of the fields
inside and outside the black hole, we have
\beq
||I_{out}\otimes\Lambda_{in} \rho_{out,in}-\rho_{out} \otimes\frac{I_{in}}{\log{d}}|| 
\leq \epsilon
\eeq
for all $\rho_{out,in}$ and some very small $\epsilon$, i.e.  we may
want to allow for some small deviations from this the map being completely randomizing.
Note that this condition of completely randomizing is stronger than demanding that
the map be merely randomizing. i.e. that
\beq
\Lambda_{in} \rho_{in}=\frac{I_{in}}{\log{d}} 
\eeq

 We should
also exclude information stored in global charges. This is because quantities
like the mass, electric charge and angular momentum are measurable outside the black
hole and are conserved during evaporation.  Amusingly, it is easy to 
use relational Lindblad operators to destroy all information 
except the information that is stored in global charges, because the Lindblad
operators commute with these global charges.
It is only the relational information one wishes to destroy.

In the case where the Hamiltonian is zero, one can
show that for virtually complete
information destruction, Lindblad operators of dimension $2 \log d$ 
are necessary and sufficient, where
$d$ is the size of the total Hilbert space (for simplicity we use a finite dimensional
Hilbert space)
We shall not reproduce the proof
here, but it is an adaptation of the proof in \cite{HaydenLSW03-approxrandom}. 
An example would be to have
Lindblad operators which are projectors onto integrals of a complete set of local projectors, and their conjugates e.g. projectors onto values of $\phi^n(x)\pi^m(x)$.
If we demand that the map be only randomizing and not necessarily
completely randomizing, then just over $\log{d}$ projectors are needed
but they should be chosen at random.  
  For Hamiltonians which are sufficiently strong or mixing, one
can use less Lindblad operators for achieving information destruction.  
This is because in the interaction
picture, the Hamiltonian evolution effectively acts a different Lindblad operator at
each sufficiently large time interval.  Understanding the time scale over which
a particular Hamiltonian and set of Lindblad operators can achieve information destruction
is a potentially interesting problem for future research.

{\bf (5) Cosmological implications}
\label{ss:horizon}
It is usually said that correlations in the cosmic microwave background (CMB)
indicate that distant parts of the universe must have been in causal contact
at some earlier epoch.  
Since the CMB is correlated over distances larger than the horizon, it is generally
believed that
the universe must have undergone some form of inflation, so that distant parts
of the universe where in causal contact in the early universe. What has not
been noted, is that correlations over distances which are space-like separated,
do not necessarily imply acuasality.  Indeed, as explained in Section
\ref{sec:causality}, non-local signaling and non-local correlations are distinct,
and the latter does not lead to a break down in causality. Correlations in the CMB
need not indicate that a signal must have propagated between the correlated regions,
or that the regions were ever in causal contact.  One can have the creation of
correlations in the CMB over space like distances in theories which do not allow 
signals to travel faster than light.
Indeed, the theories  we have examined generically create correlations over space-like
separated regions,
and we give a model theory in Appendix \ref{ss:decaycormodel} which
does exactly that.  Even more intriguing, the correlations are created over all 
length scales equally.  While this effect is 
usually small, it would be interesting to see whether it could be made to account 
for the observed CMB spectrum, in a model without inflation.

\section{Stochastic collapse models}
\label{sec:collapse}

In this paper, our primary concern has been in decoherence models and information destruction.  However, the exact
same considerations also apply to spontaneous collapse 
models\cite{grw85,grw86,pearle-csl1,gisin-percival,collapse-review}.
Such models are attempts to explain the collapse of the wavefunction, not epistemically, but 
as an actual dynamical process.  It is not our intention here to advocate for such models.  Rather, we merely wish
to point out that some of the difficulties they suffer from are of a similar nature
to the problems which plague decoherence models, and these problems 
may be addressed using the exact same
methods outlined in the previous sections.

In the classic GRW model \cite{grw85,grw86}, there is a probability per unit time that each particle is {\it hit}, meaning that
it has a probability of being localized to a Gaussian wave packet
at a particular point in space.  The more particles at a particular point,
the more likely it is to be hit, thus macroscopic states, with a greater number of particles, 
are more likely to be localized.  This leads to a possible
explanation of why large objects behave classically (in the sense that they have a well defined position).  One
of the problems
with the theory, is that it doesn't obey conservation rules.
Other problems with the theory include a difficulty in constructing a field-theoretic version, since the
non-conservation of energy leads to an instability of the vacuum. There has also yet
to be a Lorentz invariant model. Finally, as noted briefly in Section \ref{sec:qmmomentum},
the theory violates causality, on the scale
of the size of the Gaussian wave-packet to which one collapses to.

These problems
can be solved using our relational approach.  
We will briefly discuss another difficulty -- the so-called {\it problem of tails}.
We shall not go into great depth here -- we will simply present various collapse theories 
and then show how to modify them to address the problems of conservation law
violation and lack of a field-theoretic model.  
Each collapse theory can be thought of as a different way to {\it unravel the Lindblad equation}.
I.e. one has a theory which describes not just the density matrix, but each realization of the evolution in terms
of a pure state which remains pure throughout the evolution.

In the GRW \cite{grw85,grw86} model, if a hit occurs on the $i$'th particle at point $\bar{x}$, 
the wave function is multiplied by a Gaussian function
\beq
G(\bar{q}_i,\bar{x})=K\exp\left(-\frac{1}{2a^2}(\bar{q}_i-\bar{x})^2\right)
\label{eq:gaussianhit}
\eeq
where $a$ is some localization size, which can be taken $a\approx 10^{-15}$ cm,
and $q_i$ is the position of the $i$'th particle.  The probability
$p_i(\bar{x})$ that a particle is hit at point $\bar{x}$ is taken to be 
$|\bra{\psi_i(\bar{x})} \psi_i(\bar{x})\ra|^2$.
The hitting occurs at randomly distributed times according to a Poisson distribution.

We can conserve energy and momentum by not collapsing to location in terms of some absolute and unobservable
external space-time, but rather in terms of a physical meaningful relational position.  We also take the
distribution in hitting times, to be given by a distribution in terms of some physical time $\tau$ rather
than an unobservable absolute time $t$.

For example, consider the relational projector $Q_{ij}$ as in Equation 
(\ref{eq:coincidence}) acting on the i'th and j'th particle
\beq
Q_{ij}=\int d\bar{x} \ket{\bar{x}(t)}_i{}_i\bra{\bar{x}(t)}\otimes 
\ket{\bar{x}(t)}_j{}_j\bra{\bar{x}(t)}
\label{eq:hitqstate}
\eeq
With probability $|\bra{\psi_{ij}} Q_{ij}|\psi_i\ra|^2$
the $i$'th and $j$'th particle are
{\it hit}, and well end up in the state given by Equation (\ref{eq:hitqstate})
appropriately normalized.
  
To additionally conserve energy one can use the operator
\beq
\bQ_{ij}=\int d\bar{x} dt \ket{\bar{x}(t)}_i{}_i\bra{\bar{x}(t)}\otimes 
\ket{\bar{x}(t)}_j{}_j\bra{\bar{x}(t)}\otimes\ket{\tau(t)}\bra{\tau(t)}
\eeq
and the frequency of a particle being hit should be fixed in terms of the physical time
$\tau$ rather than the unobservable parameter $t$.

The fact that the wave function has a probability of taking on values of $x$ away from its peak
is known as the {\it problem of tails}. I.e. the wave-function is hit by a sharply
peaked Gaussian as in Equation (\ref{eq:gaussianhit}) and not
a delta-function.  The reason a Gaussian is chosen is that delta-function would lead to arbitrarily large
deviations from energy and momentum conservation.  This is particularly problematic in a field
theory, where such an effect makes the vacuum unstable.

Using relational observables may make the problem of tails less severe, since one is able
to conserve momentum and energy even if the wavefunction is hit by a delta function.  In such
a case, one would still have large transfers of momentum from one particle to another,
so one may still run into difficulties with experimental constraints, although often
particles that are close together are held together by some potential.
One can mitigate the energy and momentum transfer in a field theory, by using sufficiently
gentle projection operators, as described in Section \ref{sec:relationalfields}.
The problem of tails is most sever in field theories, since collapse to
a delta-function would make the vacuum unstable.  As we saw in 
Section \ref{sec:relationalfields} and \ref{sec:energy}, in
the field theory, dynamical collapse models were
stable in a field theory which has non-Gaussian collapses. 

If we still want to collapse the particles to Gaussians, then we could
hit the state with 
\beq
G(\bar{q}_i,\bar{q}_j)=K\exp(-\frac{1}{2a^2}(\bar{q}_i-\bar{q}_j)^2
\eeq
 which serves to localize particles,
but around another particle which acts as a reference frame, 
rather than around a point in absolute space.
Such a theory is slightly non-local, with the degree of non-locality being
on the scale of $a$.

Let us now turn to another model, known as 
{\it quantum state diffusion} \cite{gisin-percival}.
In such a theory, the evolution of the wavefunction is made to satisfy the Ito equation
\beq
\ket{d\psi}=-i H\ket{\psi}dt 
+ 
\12 \sum_j(2\la \gL^\dagger_j\ra \gL_j 
- 
\gL_j^\dagger \gL_j 
- 
\la \gL_j^\dagger\ra\la \gL_j \ra ) \ket{\psi} dt
+
\sum_j (\gL_j - \la \gL_j\ra) \ket{\psi}d\xi_j(t)
\label{eq:ito}
\eeq
where $d\xi_j(t)$ are independent complex and 
differential random variables representing a Wiener process.  
I.e. the mean of $d\xi_j(t)$, denoted by $\mathbb{E}(d\xi_j(t))$ 
satisfied $\mathbb{E}(d\xi_j(t))=0$ while the mean of 
$d\xi_j(t)d\xi_k(t)$ satisfies
$\mathbb{E}[d\xi_j(t)d\xi_k(t)]=\delta_{jk}dt$.
This gives the evolution of a particular wavefunction, thus it is used for dynamical collapse
models.  When averaged over the random variable however, it simply gives the diagonal form of
the Lindblad equation (\ref{eq:lindblad}).  

If we wish to have the Ito equation preserve conservation laws,
we simply act as before, constructing relational operators $Q_i$ 
via Equation (\ref{eq:genlinop}) and
using them in place of the non-relational operators $\gL_i$. If we wish
to conserve energy, we additional replace the parameter $dt$ and $t$
by a dynamical observable $d\tau$ and $\tau$.
For conservation of energy and momentum, and Hermitian $\bQ$ we would have
\beq
\ket{d\psi}=-iH\ket{\psi}d\tau
-
\12 \sum_j(\bQ_j-\la \bQ_j\ra)^2 \ket{\psi} d\tau
+
\sum_j (\bQ_j - \la \bQ_j\ra) \ket{\psi}d\xi_j(\tau)
\label{eq:itocon}
\eeq

Another dynamical collapse model is Pearle's continuous spontaneous localization model (CSL) \cite{pearle-csl1} which
modifies the Schroedinger equation as
\beq
\frac{d\ket{\psi,t}_w}{dt}=-iH\ket{\psi,t}_w-\frac{1}{4\lambda}\int d\bar{x}[w(\bar{x},t)-2\lambda A(\bar{x})]^2]
\ket{\psi,t}
\label{eq:csl}
\eeq
where $A(\bar{x})$ could be any operator, but is usually taken to be 
the number of particles in a volume $a\approx 10^{-15}$cm centered around $\bar{x}$
\beq
A(\bar{x})\equiv \frac{1}{(\pi a^2)^{3/4}}\int d\bar{z} N(\bar{z}) \exp(-\frac{(\bar{x}-\bar{z})^2}{2a^2})
\eeq
with $N(\bar{z})$ the local number operator, 
$w(\bar{x},t)$ a fluctuating field with probability density
functional
\beq
P_T(w)={}_w\bra{\psi,T}\psi,T\ra_w
\eeq
and $0\leq t\leq T$.
The resulting wave function is then
\beq
\ket{\psi,T}={\cal T}\exp(-\frac{1}{4\lambda}\int_{T_o}^{T} d\bar{x}dt[w(\bar{x},t)-2\lambda A(\bar{x})]^2])
\eeq
i.e. for a particular realization of the fluctuating field, 
it is {\it hit} by a random Gaussian centered around a particular value of $A(\bar{x},t)$.

An attempt at a relativistic model was made in \cite{pearle-relativistic}.  As the author noted, it 
was unstable due to lack of energy conservation.  
  We can make the theory stable and respect conservation laws if we replace in Equation (\ref{eq:csl}) the
operator $A(\bar{x},t)$ with our relational $Q$ or $\bQ$, and $t$ with the dynamical variable $\tau$ to
get a continuous localization model which respects conservation laws e.g.
\beq
\frac{d\ket{\psi,\tau}_w}{d\tau}=-iH\ket{\psi,\tau}_w-\sum_i\frac{1}{4\lambda_i}[w_i(\tau)-2\lambda_i \bQ_i]^2\ket{\psi,\tau}
\eeq
Furthermore,
since energy and momentum are conserved, one also may be free to have a sharper collapse
\beq
\frac{d\ket{\psi,\tau}_w}{d\tau}=-iH\ket{\psi,\tau}_w-\sum_i\frac{1}{4\lambda_i}g(w(\tau),\bQ_i)\ket{\psi,\tau}
\eeq
where $g$ leads to something other than a Gaussian (e.g. it could be the logarithm of a delta function).  Furthermore, we saw in our discussion in Section \ref{sec:lorentz},
that we can make the theory respect a form of
Lorentz invariance.  One approach is to make $\Lambda_i$ a dynamical variable
which transforms as $\partial_t$; the theory then respects minimal Lorentz invariance.

Although the above model would solve the problem of tails, it is unclear whether it is
within experimental constraints.
Note that here, $\ket{\psi,\tau}_w$ could be a field wave-functional, thus one can have a field theoretic version
of CSL.  To minimize back reaction, we could proceed as in Section \ref{sec:relationalfields}
and take the local
Lindblad operators $L_i(x)$ to be smeared local projectors e.g.
\beq
L_i(x)=\int dl g_i(l) P_l(x)
\eeq 
where the $P_l(x)$ are projection operators onto values of some local field 
field or an operator acting on the local field (such as powers of the Hamiltonian)
suitable smeared over a local test function.
The $g_i(l)$ is a Gaussian or other distribution peaked around some value of the field.  One then group-averages
this operator to construct the $Q_i$ used in any of the collapse models (while some use
only one $Q$, there is no obstacle for using many).

We have not said much about what observable the $L_i(x)$ should represent.  Indeed there
is not a good reason to a priori prefer any particular one.  Traditionally they
have been taken to be the positions of particles, but they could also be energy density,
or some other such observable.  For example something along the lines
of integrals of the local number operator or energy density, and powers of it, as discussed in 
Appendix \ref{sec:models}, are tempting choices since the strength of decoherence
in the Lindblad equation is proportional to the number of particles, and thus produces
the desired effect that macroscopic objects are more likely to be effected 
(c.f. \cite{karolyhazy66,diosi92}).

Finally, we note that although here
we have discussed the set of relational observables in the context of spontaneous
collapse models, the discussion applies to other interpretations of quantum mechanics.
For example, in Bohmian theory, the particle trajectories do not conserve energy or
momentum, and similar considerations might be applied there in order to find
the physically relevant trajectories which do conserve energy or momentum. 
Even in a many-worlds interpretation of quantum mechanics, one might consider our
relational observables as describing the relevant observables 
in a particular branch that is observed.

\section{Conclusion}
\label{sec:conclusion}

We have seen that it is possible to construct a fundamental theory which allows
for information destruction, while still being causal and respecting conservation laws.
Whether such
a theory is consistent with experiment, or is in fact realized is an open question.
However, the mere possibility of such a theory is enough to justify future study,
especially in light of the motivation coming from black hole evaporation. 
This leaves many open questions, some of which we shall turn to after briefly summarizing
our results.  

We followed a natural generalization of unitary theories to non-unitary theories, through
the Lindblad equation of (\ref{eq:lindblad})
\beq
\frac{d\rho}{dt}=-i[H,\rho]-\12\sum_{ij}\gamma_{ij}(\gL_j^\dagger \gL_i \rho 
+ \rho \gL_j^\dagger \gL_i -2\gL_i \rho \gL_j^\dagger) \s .
\nonumber
\eeq
This is the most general evolution one can have which is a completely positive map
(doesn't change the meaning of local density matrices), and is Markovian.
However, in order to ensure that conservation laws were respected, we replaced 
the Lindblad operators $\gL_i$ with the relational operators
\begin{equation}
Q_i=\int  dg U(g)\gL^{(1,2,3...)}_iU^\dagger(g)
\nonumber
\end{equation}
where we average over some unitary representation of a group so that our evolution
conserves the generator of that group.  This allows us to respect the traditional 
conservation laws. We also introduced 
extra fields $1,2,3...$ which allows our operator $Q_i$ to be nontrivial, as it can
describe relations between different fields.  We argued in Section \ref{ss:richness} 
that such operators allow for the destruction of the information concerning 
the sorts of observables we make in the lab.

Energy conservation needed to be treated slightly differently.
The first step is the same as above, and involved group averaging, using the Hamiltonian
as a generator of time translations.  The resulting Lindblad operators conserved energy
but were no longer local.  In order to make them local, we imposed time-translation invariance
on the state,
\beq
\frac{d\rho}{dt}=0
\nonumber
\eeq
which yielded an equation akin to the Wheeler deWitt equation.  This equation was local
and conserved energy, and was non-trivial in the sense that one could describe evolution
in terms of correlations between observables and a clock.

In Section \ref{sec:causality}, we showed that this evolution respected causality.  
This was done by taking our relational Lindblad operators to be integrals of local
operators, and taken them to be either Hermitian, or to come in a pair with their
Hermitian conjugate.
We also
distinguished another effect -- the creation or destruction of spatially 
separated correlations.  This an effect was not seen in the non-relativistic limit,
but was seen for relativistic fields.  Fortunately, it does not result in a break-down
of causality.  Certainly this evolution of correlations is non-intuitive, but it cannot
be used to signal superluminally.  We discuss the effect of correlations in a model given
in Appendix \ref{ss:decaycormodel}.

In Section \ref{sec:noether}, we also took on the question of to what extent any symmetry should imply a conservation law. We derived a generalization of Noether's theorem and
the continuity equation for Markovian non-unitary evolution.
This placed a restriction on the Lindblad operators -- they either had to be Hermitian (in which
case, the symmetry implied the conservation law), or they had to be creation or annihilation
operators.  Combined with the locality constraints, this appears to restrict the violation
of conservation laws to be mild.  It remains an open question whether further restrictions
exist which would require exact conservation from the symmetry, or whether such a restriction
ought to be imposed to keep the theory finite.

In Section \ref{sec:lorentz} we discussed the fact that our theories were invariant under
the orthochronous Lorentz group, but proved that it was impossible to get Lindblad evolution
where the dissipater term transforms like a four-vector.  We also show in Appendix 
\ref{sec:retrodiction}, that the theory could be made time-symmetric.

Finally, in Section \ref{sec:collapse},
 we discussed application of the relational approach taken here, to stochastic
collapse models believed by some to describe the emergence of classicality
from quantum mechanics. While in Section \ref{sec:coupletogravity} we discussed toy models
for coupling the theory to gravity.

We have left open many theoretical questions, and we take the opportunity in
this conclusion to summarize some of them.
From an experimental point of view, one might test some of the models, by looking 
for momentum exchange between pairs of particles.  Such an effect may lead
to spontaneous ionization of atoms, as an electron and nucleus are given a kick
in the opposite direction\cite{squires-ionization}. 
The existence of such
an effect, and it's magnitude, depends on
which model is chosen.  In some models, the effect can be made very small making it unlikely
that current experimental constraints could rule out the theories considered here,
although they may restrict the parameters in some models.
Likewise, long range creation of correlations, or abnormally fast decay of correlations
is a signature of all the models, but we suspect its effect would be very difficult to measure.

From a theoretical point of view, some of the more pressing open questions concern
the richness of local relational observables, as noted in Section \ref{ss:richness}.
To what extent can such observables describe the world around us, or destroy 
the required information in a black hole.  To explore this, it may 
be interesting to look at numerical simulations
of some of the models discussed in Sections \ref{ss:richness}
and Appendix \ref{sec:models}, particularly those which result in effective
potentials which depend on distance.  

The constraints due to causality discussed
in Section \ref{sec:causality} and Appendix \ref{ss:superluminal} 
were not fully explored, and more work is needed in
determining which Lindblad operators are consistent with causality, and lead to
renormalisable theories.  Even constructing additional models to the one we described
would be useful. 

In terms of coupling the theory to gravity, we only mentioned possible toy models, and
it would be interesting to pursue this further.  Especially with respect to theories
which don't conserve the energy-momentum tensor, since the difficulty in coupling such
theories to gravity might preclude such theories.
With regard to gravity, we noted that correlations due to fundamental decoherence could
be imprinted in the cosmic microwave background, and it would be interesting to explore this
further, possibly as an alternative to inflation.

With regard to the generalization of Noether's theorem proven in Section \ref{sec:noether}
it would be interesting to see whether it could be applied to any condensed matter systems.
Nothing about it depends on whether the non-unitarity is fundamental, we just require
that some symmetry be respected, perhaps only approximately.

\acknowledgments{We would like to thank Yakir Aharonov, 
Robert Alicki, Daniel Gottesman, Noah Linden,
 Fotini Markopoulou, Philip Pearle, David Poulin, John Preskill, Rob Spekkens, 
Lenny Susskind, David Tong, and Bill Unruh for interesting discussions. We are 
especially grateful for discussions with Sandu Popescu over the years, which
have greatly influenced this work.  Much of this research was presented at the 
Quantum Information in Quantum Gravity conference at the Perimeter Institute,  
December 10th, 2007. J.O. is supported by the Royal Society.
B.R. would like to acknowledge the Israel science foundation
grant 784/06 and the German-Israeli foundation
grant I-857.
}
\bibliographystyle{jhep}
\providecommand{\href}[2]{#2}\begingroup\raggedright\endgroup

\appendix
\section{Models}
\label{sec:models}

In the body of this paper we have tried to keep our considerations as
general as possible, using particular models only for illustrative purpose. In
this appendix we will briefly sketch a few particular models, which we have either
already used, or which we have found useful.  We will not explore them in any
detail, but rather hope that they might prove useful for future research.

We will first introduce Lindblad operators based on a local number operator.
The motivation for this is that it is a field theoretic model, whose non-relativistic limit
gives the quantum mechanical model of Equations (\ref{eq:coincidence}-\ref{eq:lindbladqmm}).
%
%
We wish to construct a field theoretic model which has as its limit,
the quantum mechanics model of Equations (\ref{eq:coincidence}-\ref{eq:lindbladqmm}).
To this end,  consider a scalar field and define
\beq
\psi(x)\equiv\phi(x) + i \pi(x) \label{eq:localfield} 
\eeq 
In the non-relativistic limit $\psi^\dagger(x)$ acts as a 
 localized creation operator i.e. it creates a single particle which is localized
around the point $x$
\beq 
|1_x\ra =  \psi^\dagger(x)|0\ra 
\eeq

We wish to construct a field theoretic model which has as its limit,
the quantum mechanics model of Equations (\ref{eq:coincidence}-\ref{eq:lindbladqmm}).
To this end,  consider a massive scalar field and define
\beq
\psi(x)\equiv\phi(x) + i \pi(x) \label{eq:localfield}
\eeq
The operator   $\psi^\dagger(x)$ acting on the vacuum creates a single particle state
\beq
|1_x\ra =  \psi^\dagger(x)|0\ra
\eeq
which is localized around the point $x$.

For modes with $k\ll m$ we get $\psi(x) \propto \int dk e^{ikx} a_k$ and hence for non-relativistic modes $\psi$ and $\psi^\dagger$ acts as
a non-relativisitic annihilation  operators.

One can verify that $|1_x\ra$ transforms correctly under translations
i.e.
 \beq U_l \proj{1_x} U_l^\dagger = \proj{1_{x+l}}
\eeq
where
$U_l=\exp(i l P)$ and $P$ is the field momentum operator.

Furthermore, one also has
\beq
\la 1_x |
1_{x+l}\ra = {1\over 2\pi} \int dk (2+ \omega_k + {1\over \omega_k}) e^{ik l} \to 0
\eeq
for $l>>>1/m$. In this case there is a region for which $\omega_k + {1\over \omega_k}\approx
constant$ and the integral tends to zero.
Hence the state $|1_x\ra$
indeed describes a single particle state that is localized around $x$ over a region
of size $~1/m$.

We can then define a  number operator as $N(x)=\psi^\dagger(x)\psi(x)$ with $\psi$
defined as in Eq. (\ref{eq:localfield}).  Now imagine two fields, so that we
have $N_1(x)$ and $N_2(x)$.  Then take
 \beq
Q=\int dx N_1(x)\otimes N_2(x)
\label{eq:twonumber}
\eeq
as the Lindblad operator we are interested in.  In the single particle non-relativistic limit, i.e. the
probability of getting $N_1(x)\geq 1$ or $N_2(x)\geq 1$  is very small, this
Lindblad operator clearly reduces
to that of Equation (\ref{eq:coincidence})
\beqa
Q=\int dx \proj{x}\otimes\proj{x}
\nonumber
\eeqa
We thus see that we recover the quantum mechanical model in the non-relativistic limit.


If we wish to have a field act as a reference frame, then another method is
to use coherent states.  In this limit, we can have a fully field theoretic
model which can reproduce the models
of Equation (\ref{eq:fieldtoparticle}) used to show that one could decohere
a field at a point and still conserve momentum.
The coherent states behave in some sense like a classical system, 
and thus behave like localized particles.
Recall that for a harmonic oscillator we can construct a coherent state
\beq
|g\rangle :=e^{g a + g^* a^\dagger} |0\rangle_{HO}
\eeq
from $|0\rangle_{HO}$ the ground state of the Harmonic oscillator 
such that $a|g\rangle = g|g\rangle$, $\langle g|a^\dagger= g^*\langle g|$.

Since each field mode is a harmonic oscillator we can construct 
something like $|\Psi_{coh}\rangle=\otimes_k |g_k(x_0)\rangle $ for the field
\beq
|\Psi_{coh}\rangle=e^{\int dk(g(k) a_k + g^*(k) a_k^\dagger)}|0\rangle
\eeq
If we compute the expectation value of the field at a point it will be
\beq
\langle\Psi|\phi(x)|\Psi\rangle = \int {dk\over 2\pi\omega_k} ( g(k)e^{ikx}
+g^*(k) e^{-ikx})
\eeq
so now $a_k$ is an eigenstate and the expectation value can be 
chosen by taking suitable $g(k,x_0)$ to be concentrated around $x_0$,
so that $\langle \phi(x)\rangle = g(x-x_0)$.

Suppose we have two fields, and choose 
\beq
Q = \int dx \phi_1(x) \phi_2(x)
\eeq
although any local operator $L(x)$ can be used in place of $\phi_2(x)$.  Let us
look at the dissipater term
\beq
{\cal D}(\rho)=-\gamma\frac{1}{2}[ Q,[Q,\rho]
\eeq
if we assume that $\rho \simeq \rho_1\otimes\rho_2$ and take e.g. 
$\rho_2= |\Psi_{coh}\rangle\langle \Psi_{coh}|$
than under the assumption that the evolution still keeps 
$\rho_2$ close to the chosen coherent state
the dissipater term is roughly give by
\beq
{\cal D}(\rho)
\approx
-\gamma \frac{1}{2}\int dx dx' g(x-x_0)g(x'-x_0)  [\phi_1(x)\phi_1(x'),[\phi_1(x)\phi_1(x'),\rho]]
\eeq
which  is exponentially small if 
x and x' are not close to $x_0$. 

If it is possible to take the limit that $g(x-x_0)$ are delta 
functions we get that effectively  the dissipater term is 
\beq
{\cal D}(\rho)
\approx-\frac{1}{2}\gamma [\phi^2(x_0),[\phi^2(x_0),\rho]]
\eeq
a decoherence term at a point, and yet we did not violate momentum conservation.

Turning back to models which use number operators
such as those of Equation (\ref{eq:twonumber}), we might generalize this
to consider higher and higher powers of $N(x)$, i.e.
\beq
Q_{mn}=\int dx N_1^m(x)\otimes N_2^n(x) 
\eeq
in order to get at the full richness of the theory.  

In such a case, one can even
consider only one field, since such a theory can be made {\it self-referential} i.e. if we have
two particles and only one species, then the set of Lindblad operators
\beq
Q_m=\int dx N^m(x)
\eeq
will distinguish between the cases where the particles are coincident or not.  In the limit
where we only have two localized particles, one only needs $Q_2$, 
since non-coincidence gives $Q_2=2$, while
coincidence gives $Q_2=4$.  In effect, one has a know background field $\phi_o$
which acts as a reference frame, and one treats the remainder, $\delta\phi=\phi-\phi_o$
as a perturbation which is being measured or decohered.  Thus to first order
\beq
Q_2 \approx N_o + 2 \int dx N_o(x)\delta N(x)
\eeq
which, when $N_o$ is fixed, acts very similar to the two field model described 
by Equation (\ref{eq:twonumber})

It is also very tempting to consider models which have Lindblad operators
as integrals of powers of the Hamiltonian
\beq
{\bar Q}_m=\int {\cal H}^m(x) d{\bar x}dt
\eeq
However, it remains to show that such a theory is renormalisable.  Nonetheless, such
models may be particularly interesting for stochastic collapse theories, since
the greater the energy density, the greater the probability of collapse.

To avoid back-reaction, and keep the theory finite, we might want to smear over 
field observables.  This can be done both over a small region of space, and
also over values of the field.  We discussed this at the end of Section
\ref{sec:relationalfields}, and so shall not repeat that discussion here.

In Section \ref{ss:richness} we discussed why one might want to consider models
which have tiny violations of non-locality such as those given by Equation 
(\ref{eq:nonlocallop}).
\beqa
Q_{l\alpha\beta}=\int dx A_\alpha(x) B_\beta(x+l)
\nonumber
\eeqa
we might then integrate over $l$ to create a smeared Lindblad operator over some 
small range of $l$.  

An analogous construction can be considered as follows. 
Let $E_x=\proj{1_x}$ be the projection-like operator on a localized single
particle defined above, and $\tilde E_x$ be the same type of operator for a 
second field.
Now define the Lindblad operator
\beq 
Q = \int dx dx' \tilde E_x E_{x'} h(x-x') 
\eeq 
where $h(x-x')$ is a
"hat" shaped function, $h=1$ for $|x-x'|<\Delta$.

Now we have two scales $1/m$ the "size" of the localization of $E_x$, and $\Delta$ the
size of the smearing.
An interesting case is $\Delta\gg 1/m$, where we smear over scales larger then the
Compton wave length of the two fields (i.e. we require  $\Delta >>
\max (1/m, 1{\tilde m})$)

Under this assumption  $Q^2 \approx Q$. I.e. in the limit it
becomes a projector.
\beq 
Q^2=\int dx dx' dy dy' \tilde E_x \tilde E_y  E_{x'} E_{y'}h(x-x')h(y-y')
\eeq 
but since most of the integration is over regimes for which $x-x'\gg 1/\tilde
m$ and $y-y'\gg1/m$ we can replace 
\beq
\tilde E_x \tilde E_y \approx
\delta(x-y) \tilde E_x
\eeq
\beq  
E_{x'}  E_{y'} 
\approx
\delta(x'-y') \tilde E_{x'} 
\eeq
so we get 
\beq 
Q^2= \int dxdy \tilde E_x E_{x'} h^2(x-x') 
\eeq but
since $h=1$ we obtained  $Q^2=Q$ (equality of course only in the limit of
$m\Delta \to \infty$, but we can perhaps expect, that $Q^2$ tends to $Q$ up to
a correction that dies of like $\exp-(m\Delta)$)

The key point is that due to the smearing function the field which
acts as a reference frame is "too
big" to resolve close to the Compton length scale. Once we do that, we retrieve
that $Q$ is a good projector even in the fully relativistic case.
This smeared projector commutes with $P_{total}$ since the function $h(x-y)$
depends on relative coordinates.

On a completely different note, recall in Section \ref{ss:richness} 
that there are models in which
one can add potentials to discriminate between particles, or localized fields at
different distances.
I.e. the Hamiltonian terms might become very important
when one has an effective potentials $V(x-y)$ which depend on the distances between
objects and couples to other degrees of freedom (such as local spins).  This can serve to
distinguish distances between particles.  Since analyzing such models probably
requires numerical calculations, we will not treat such models in more detail here.

Finally, another set of possibilities, is to take operators which are {\it coherently group-averaged}
as an example, first recall that  our coincidence 
Lindblad operator of Equation (\ref{eq:coincidence}) was 
\beqa
Q=\int d\bar{x} \proj{\bar{x}}\otimes\proj{\bar{x}}
\nonumber
\eeqa
  which is an incoherent mixture of coincidences, representing the fact that one
  lacks knowledge of the absolute reference frame $\bar{x}$.  However, one could
  consider the function generated by coherent group-averaging
  \beq
  \ket{\psi_k}=\int d\bar{x} e^{-ik\bar{x}}\ket{\bar{x}}\ket{\bar{x}}
  \eeq
  Just as before, this commutes with $P$ (it is an eigenstate of $P$, rather
  than a mixture of eigenstates. The projectors $Q_k$ onto $\ket{\psi_k}$ could then
  be used in the Lindblad equation.  These operators are rather trivial,
  but this appears to be a function of the simplicity of the case being considered.

For the field, one can consider something similar, by looking 
at operators build out of a coherently group-averaged operator
\beq
{\tilde M_i}=\int dg dg' U(g) M_i U^\dagger(g')
\eeq
This operator is invariant under incoherent group averaging, and will thus conserve the required
quantity.  However, it is not always local.

\section{Correlations in a particular model}
\label{ss:decaycormodel}

Here we examine a particular model and show that 
the creation or destruction of correlations occurs with very small probability
on typical states.  I.e. it takes place on a vanishingly small part of the
full state space.
For simplicity, we will consider a quantum mechanical version of the field theory operator of Equation (\ref{eq:positionprojs}), however, our considerations
will apply to a field theory.  To this end, let us take $\proi{x}$ to be 
a set of smeared projectors (one might be inclined to imagine that they
are akin to projections onto the values of 
some field observable $\vp(x)$) in a discretization of $\vp(x)$).  
Let us take as an example, a model in which we can write
\beq
\Q{i}=\int dz \proi{z}
\label{eq:positionprojs}
\eeq
with $\proi{z}$ some projector or smeared projector.
Next, consider only two points
in space, $x$ and $y$, and 
define
\beq
\pplus=\proi{x}\ot I(y) +I(x)\ot \proi{y} \s .
\label{eq:pplusprojector}
\eeq

Then dropping
the Hamiltonian term for simplicity we get that evolution of an
operator $A(x)$ is local, and of Lindblad form.  I.e.
\beq
\frac{d}{dt}\A{}=-\hi [\proi{x},[\proi{x},A(x)]]
\label{eq:localmodelevolution}
\eeq
as we expect from Section \ref{sec:causality}.
However, the correlation $\A{}\ot \B{}$ of two distant observables 
behave as
\beqa
\frac{d}{dt}[\A{}\ot \B{}]
&=&-\frac{1}{2}\sum_{i}\hi [\Q{i},[\Q{i},[\A\ot\B]]
\nonumber\\
&=&-\12\sum_{i} \int dz dz' \hi
[\proi{z},\proi{z'},[\A\ot\B]]
\nonumber\\
&=&
-\12 \sum_{i} \hi
[\pplus,[\pplus, \A\ot\B]]
\label{eq:lindpplus}\\
&=&
\frac{d}{dt}\A{}\ot\B{}
+
\A{}\ot\frac{d}{dt}\B{} + V(\A\ot\B)
\eeqa
where $V$ is the term which violates the product evolution form
of Equation (\ref{eq:product}), and is equal to
\beq
V=-\sum_{i} [\proi{x},\A{}]\ot [\proi{y},\B{}]
\label{eq:prodviolprojm}
\eeq
this violating term looks a bit like a Lindblad form, as it can be written as
\beq
V=-\sum [\proi{x},[\proi{y},\A\ot\B]]
\eeq
although the matrix of projectors is not positive.

This terms is zero on the subspace where $\vp(x)\neq \vp(y)$.  I.e. consider the projector
$P_{dis}$ onto this subspace 
\beq
P_{dis}=\sum_{i\neq k,j\neq l}\pai{x}{i}\pbi{x}{j}\pai{y}{k}\pbi{y}{l}
\eeq
then
\beq
V(P_{dis} \A\ot\B P_{dis})=0
\eeq

 For a finite dimensional Hilbert space of dimension
$n$ at each point in space, the subspace in which $V$ is non-zero 
grows as $n$ while the total space grows as $n^2$.  Thus as $n$ goes to
infinity, the fraction of the Hilbert space in which one finds creation or destruction of correlations goes to
zero.
Thus for typical states, destruction or creation of correlations will be a very small effect. 
If we increase the number of fields and how fine grained the projectors are, this term becomes rarer still.

It is worthwhile to see the effect of the term $V(\rho_{AB})$ on the evolution of a state and how it violates
the product evolution form of Equation (\ref{eq:product}).  Let us consider a finite dimensional system
and density matrix at points $x$ and $y$ decomposed in terms of the basis $\proj{i_x}=\proi{x}$, 
\beq
\rho=\sigma_{ij,kl}\ket{i_x}\bra{j_x}\otimes\ket{k_y}\bra{l_y}
\label{eq:abstate}
\eeq
The solution of Equation (\ref{eq:lindpplus}) with and without the product violating term $V$ is
given in Table \ref{tab:correlations} for $\hi=\gamma$.  Note that the evolution is only non-product for
terms $\ket{i_x}\bra{j_x}\otimes\ket{i_y}\bra{j_y}$.  These off-diagonal elements decay faster than they would
under purely product decoherence which obeys Equation (\ref{eq:product}).
\begin{table}
\label{tab:correlations}
  \begin{tabular}{ | l || c | r | }

   \hline
condition on $i,j,j,l$ &   with $V(\sigma)$ & without $V(\sigma)$  \\ \hline\hline
& $\sigma_{ii,kk}(t)=\sigma_{ii,kk}(0)$ & $\sigma_{ii,kk}(t)= \sigma_{ii,kk}(0)$ \\ \hline
$ k\neq l$ & $\sigma_{ii,kl}(t)=   e^{-\gamma t}\sigma_{ii,kl}(0)$ &  
$ \sigma_{ii,kl}(t)= e^{-\gamma t}\sigma_{ii,kl}(0)$ \\ \hline
& $\sigma_{ij,kk}(t)=   e^{-\gamma t}\sigma_{ij,kk}(0)$ &  $ \sigma_{ij,kk}(t)= e^{-\gamma t}\sigma_{ij,kk}(0)$\\ \hline
$k\neq l, i\neq j, i\neq k, j\neq l$ &$ \sigma_{ij,kl}(t)=   e^{-2\gamma t}\sigma_{ij,kl}(0) $
       &   $\sigma_{ij,kl}(t)= e^{-2\gamma t}\sigma_{ij,kl}(0)$\\ \hline
& $\sigma_{ij,ij}(t)=   e^{-4\gamma t}\sigma_{ij,ij}(0)$ &  $ \sigma_{ij,ij}(t)= e^{-2\gamma t}\sigma_{ij,ij}(0)$ \\ 
    \hline
    \end{tabular}
\caption{Evolution of correlations of the density matrix 
$\rho=\sigma_{ij,kl}\ket{i_x}\bra{j_x}\otimes\ket{k_y}\bra{l_y}$ 
with and without the product violating
term $V(\sigma)$}
  \end{table}

This shows how correlations can decay faster than one might otherwise expect.
One also finds creation of correlations.  As an example, take as an initial state
that of Equation (\ref{eq:abstate}) and with all $\sigma_{ij,kl}$ equal.  
This is an initial
uncorrelated pure state, with each party's state in a superposition over all basis states,
i.e. we will take the state at $x$ to be
$\ket{\psi_A}=\sum \ket{i}/\sqrt{n}$ and similarly for the state 
$\ket{\psi_B}$ at $y$.  
Now, we know from  
the local evolution law of 
Equation (\ref{eq:localmodelevolution}) that all off-diagonal terms
of the local states will decay, and the local states will be maximally mixed.
I.e. the state at $x$ will evolve to $\rho_A=\sum \proj{i}/n$, and similarly for
$\rho_B$ at $y$.  The local entropy of each state is thus $\log n$.
However, from Equation (\ref{eq:pplusprojector}), we see
that there are only $n$ projectors $\pplus$.  Thus the total state
will only be decohered into $n$ states, rather than $n^2$ possible
states.  In particular, although each local density matrix is maximally
mixed, the total density matrix is decohered into the states 
\beq
\ket{\psi_i}=\sum_{jk}\ket{ij+ki}/\sqrt{2n}
\label{eq:degsurvive}
\eeq
which can be seen from the fact that these
are the states which survive when we apply the Lindblad operator to
both sides of the initial state
\beq
\pplus\proj{\psi_A\psi_B}\pplus=\proj{\psi_i}
\eeq

To put it another way, each Lindblad operator $\pplus$ is degenerate,
selecting states where one of the two sites is in state $\ket{i}$, but the
other site can be in any state.
Thus terms which are superpositions over these degenerate states
will survive (the states of Equation (\ref{eq:degsurvive})).  Thus although
the local states will look maximally mixed, the total state 
will contain coherences.  As a result, we have that the entropy
of each system is $n$ while the entropy of the total system is also $n$.
Since the mutual information $I(A:B)$ between two systems $A$ and $B$
is 
\beq
I(A:B)=S(A)+S(B)-S(AB)
\eeq
where $S(A)$ and $S(B)$ is the Von-Neumann entropy $S=-\tr \rho \log\rho$ of the state at $x$ and $y$ respectively
and $S(AB)$ is the entropy of the joint system.  In the present case,
the entropy of the local states is $\log n$, while the entropy
of the total state is no larger than $\log n$ (the number of states the system
is decohered into), and thus the mutual information is at least $\log n$ and
approaches it for large $n$ since then is when the states of Equation
(\ref{eq:degsurvive}) become orthogonal.

\section{A time-symmetric formulation}
\label{sec:retrodiction}


A Lindblad equation has an implicit boundary condition at $t=0$, because if
you evolve the equation backwards in time past $t=0$ the equations no longer correspond to a CPT map i.e. the density matrix will evolve into something which
is not a density matrix.  This is reasonable, since at $t=0$ we assume some
initial condition (such as the state being pure), and then evolve forward in time.

As a result, it is often said that the Lindblad equation gives
an arrow of time, since as we evolve it, a state become more mixed.
I.e. it is said that since the evolution gives an entropy increase, it must
necessarily give an arrow of time.
However, this simply reflects  a choice of time-asymmetric 
boundary conditions -- i.e. we start with a pure state at $t_0$ and find that we
don't have complete knowledge of its state at future times.  Yet, given
a pure state at $t_0$ we could
just as well have tried to retrodict what the state would be at earlier times.
In this case, if the fundamental evolution is not unitary, then we will also not have
complete knowledge of the state in the past.  Such a situation has been contemplated
for retrodicting the results of 
past measurements from future measurements, both in standard quantum
theory\cite{abl}, and in open quantum systems\cite{barnett-retrodict}.

Here, we are interested in 
constructing a full retrodicting Lindblad equation for the density matrix itself.
This has an entropy increase backwards in 
time, which is what one expects since for an initially pure state, 
you can't predict with certainty which state
led to the current pure state.   Thus we see that entropy not only increases into
the future, it also increases into the past -- therefore, it does not give an arrow of time.

The retrodiction equation can be constructed
from the forward evolving equation by 
changing the sign of the dissipater ${\cal D}(\rho)$
so that if we were to take $dt\rightarrow -dt$, we would get Lindblad evolution
backwards in time which is valid for earlier times.  
\beq
\frac{d\rho}{dt}=-i[H,\rho]+\12\sum_{ij}\gamma_{ij}(L_j^\dagger L_i \rho 
+ \rho L_j^\dagger L_i -2L_i \rho L_j^\dagger)
\label{eq:lindbladbackwrds}
\eeq
We could also derive the above equation by considering a microscopic derivation,
where we have a unitary interaction and trace out an environment. I.e. starting
from the usual forward evolving equation
\beq
\frac{d\rho_{SE}}{dt}=-i[H_E+H_S+H_{ES},\rho_{SE}]
\eeq
which leads to the usual forward evolving Lindblad equation,
we take the backward evolving unitary evolution
\beq
\frac{d\rho_{SE}}{dt}=i[H_E+H_S+H_{ES},\rho_{SE}]
\eeq
to get the backward evolving version of Equation (\ref{eq:lindbladbackwrds})
\beq
\frac{d\rho}{dt}=i[H,\rho]-\12\sum_{ij}\gamma_{ij}(L_j^\dagger L_i \rho 
+ \rho L_j^\dagger L_i -2L_i \rho L_j^\dagger)
\label{eq:lindbladbackwrds2}
\eeq
where we have removed the labels $S$ and $E$ for system and environment.
In these last two equations, a positive $dt$ represents a step back in time, so
we then take $dt\rightarrow -dt$ to ensure that time is defined in the standard way,
thus arriving at Equation (\ref{eq:lindbladbackwrds}).
We can combine the integral version of both the forward and backward
evolving Lindblad equation to define
the evolution over all times $t$ given some state at $t=0$
\[ \rho(t)=\rho(0)-i\int_0^t dt [H,\rho] + \left\{ \begin{array}{ll}
      -\int_0^t dt \12\sum_{i}\gamma_{i}(L_i^\dagger L_i \rho 
+ \rho L_i^\dagger L_i -2L_i \rho L_i^\dagger)    & \mbox{if $t \geq 0$};\\
     +\int_0^tdt \12\sum_{i}\gamma_{i}(L_i^\dagger L_i \rho 
+ \rho L_i^\dagger L_i -2L_i \rho L_i^\dagger)    & \mbox{if $t < 0$}.\end{array} \right. \]


\section{No instantaneous signalling and causality} 
\label{ss:superluminal}

In Section \ref{sec:causality}, we saw that as long as we chose the Lindblad operators
to be integrals of local operators and either Hermitian, or coming in
pairs with its Hermitian conjugate, then the evolution of of Equation (\ref{eq:lindbladq}) will be causal.
In the present section, we note a more general locality condition for the Lindblad
operators which is exact, but under the weaker requirement that 
the equations of motions of local observables are local at an instant, i.e. their
evolution only depends on the value of local fields.  
This is equivalent
to saying that there is no instantaneous signalling. 
I.e. for a local Hermitian observables $A(x)$

\beqa
\frac{dA(x)}{dt}&=& {\cal L}(A(x)) \nonumber\\
&=&
f(x)
\label{eq:causal}
\eeqa
with $f(x)$ some local operator.  This need not guarantee that signals cannot propagate
at speeds faster than light, but it is a necessary condition.

We can also write this condition as a vanishing of the equal time commutator at distant locations:
\beq
[\frac{dA(\bar{x},t)}{dt}, B(\bar{x}',t)] = C(\bar{x},t)\delta(\bar{x}-\bar{x}')
\eeq
For $dA(\bar{x},t)/dt$
we can substitute the right hand side of Equation (\ref{eq:lindbladq}), and expand the Lindblad
operators as 
\beq
Q_i=\int d\bar{x} L_i(\bar{x})
\eeq
where we make no assumption about $L_i(\bar{x})$.  This gives the general condition that for 
$x\neq x'$
\beq
\sum \gamma_{i}
\int d\bar{z} d\bar{y} 
[L_i^\dagger(\yv) [L_i(\zv),A(\xv)]+[A(\xv),L_i^\dagger(\zv)] L_i(\yv) , B(\xv')] =0
\label{eq:localitycondition}
\eeq
It comes as no surprise, given the results of the previous Section \ref{sec:causality},
that one way to satisfy this condition is to require that for any local Hermitian
operators $A(x)$ and $B(x)$: 
\begin{enumerate}
\item
\begin{enumerate}
\item $[L_i(\zv),A(\xv)]=D_{i,A}(\xv)\delta(\xv-\zv)$
\item $[D_{i,A}(\xv),B(\zv)]\propto \delta(\xv-\zv)$
\end{enumerate}
and that either
\item $L_i(\xv)=L_i^\dagger(\xv)$
or 
\item for any $\sqrt{\gamma_{i,0}}L_{i,0}(\xv)=\sqrt{\gamma_{i,1}}L_{i,1}^\dagger(\xv)$, 
\end{enumerate}
where we have used a double index for the last condition
in order to ensure that the Lindblad operators come in pairs with their
Hermitian conjugate as in Equation (\ref{eq:lindbladqfromtwofieldnonherm}).
Imposing condition 1 (a) and (b)
(i.e. that 
$L_i(\xv)$ is a local operator) on Equation 
(\ref{eq:localitycondition}) gives
\beq
\sum \gamma_i
\left(
D_{i,B}^\dagger(\xv')
D_{i,A}(\xv)
-
D_{i,B}(\xv') D^\dagger_{i,A}(\xv)
\right)
\eeq

I.e. that $\sum \gamma_i D_{i,B}^\dagger(\xv')D_{i,A}(\xv)$ is Hermitian
for all $A(x),B(x)$.
This can be satisfied by taking the $D_{i,A}(\xv)$ 
anti-Hermitian or Hermitian, which is equivalent to taking the
$L_i(\xv)$ 
Hermitian (Condition 2). It can also be satisfied by instead
taking Condition 3.


For the more general
condition of
Equation (\ref{eq:localitycondition}), it is 
not clear what other ways exist where one could
satisfy it
nor whether other conditions would violate
full causality.  We conjecture that Conditions 1 and either 2 or 3 are required.


\end{document}